\documentclass[10pt,preprint2]{aastex}
\usepackage{psfig}
\shortauthors{Perrett et al.}
\shorttitle{The M31 Globular Cluster System}

\def\ltsim{\mathrel{\hbox{\rlap{\hbox{\lower4pt\hbox{$\sim$}}}\hbox{$<$}}}}
\def\gtsim{\mathrel{\hbox{\rlap{\hbox{\lower4pt\hbox{$\sim$}}}\hbox{$>$}}}}
\def\kms{\hbox{km$\,$s$^{-1}$}}

\begin{document}
 
\title{The Kinematics and Metallicity \\ of the M31 Globular Cluster 
System}

\author{K. M. Perrett,\altaffilmark{1}
\altaffiltext{1}{Department of Physics, Queen's University, Kingston,
ON K7L 3N6, Canada}
T. J. Bridges,\altaffilmark{2}
\altaffiltext{2}{Anglo-Australian Observatory, Epping, NSW, 1710 Australia}
D. A. Hanes,\altaffilmark{1}
M. J. Irwin,\altaffilmark{3} \altaffiltext{3}{Institute of Astronomy,
Cambridge University, Cambridge CB3 0HA, England, UK}
J. P. Brodie,\altaffilmark{4} \altaffiltext{4}{UCO/Lick Observatory,
University of California at Santa Cruz, Santa Cruz, CA 95064, USA}
D. Carter,\altaffilmark{5} \altaffiltext{5}{Liverpool John Moores
University, Twelve Quays House, Egerton Wharf, Birkenhead, CH41 1LD,
England}\\
J. P. Huchra,\altaffilmark{6} \altaffiltext{6}{Harvard-Smithsonian
Center for Astrophysics, 60 Garden Street, Cambridge, MA 02138, USA}
and F. G. Watson\altaffilmark{2}} 

\begin{abstract}
With the ultimate aim of distinguishing between various models
describing the formation of galaxy halos (e.g. radial or multi-phase
collapse, random mergers), we have completed a spectroscopic study of
the globular cluster system of M31.  We present the results of deep,
intermediate-resolution, fibre-optic spectroscopy of several hundred
of the M31 globular clusters using the Wide Field Fibre Optic
Spectrograph (WYFFOS) at the William Herschel Telescope in La Palma,
Canary Islands.  These observations have yielded precise radial
velocities ($\pm 12$~\kms) and metallicities ($\pm0.26$~dex) for over
200 members of the M31 globular cluster population out to a radius of
1.5 degrees from the galaxy center.  Many of these clusters have no
previous published radial velocity or [Fe/H] estimates, and the
remainder typically represent significant improvements over earlier
determinations.  We present analyses of the spatial, kinematic and
metal abundance properties of the M31 globular clusters.  We find that
the abundance distribution of the cluster system is consistent with a
bimodal distribution with peaks at ${\rm [Fe/H]}\sim-1.4$ and $-0.5$.
The metal-rich clusters demonstrate a centrally concentrated spatial
distribution with a high rotation amplitude, although this population
does not appear significantly flattened and is consistent with a bulge
population.  The metal-poor clusters tend to be less spatially
concentrated and are also found to have a strong rotation signature.

\end{abstract}



\section{Introduction}

The mechanisms involved in galaxy formation remain one of the major
unsolved problems in astronomy.  Over the past several decades,
various groups have proposed a range of models which endeavor to
explain the observed properties of galaxies and their globular cluster
systems (GCSs).  One early formation model from \citet{egg62} argues
for a single, large-scale collapse of material to form galactic bodies
such as the Milky Way.  A principal competing model maintains that
formation has occurred through random mergers of fragmented gas clouds
over the course of billions of years, implying a hierarchical origin
\citep{sea78, col00}.  In order to interpret the bimodality seen in
the globular cluster metallicity and color distributions of many
galaxies \citep{ash98}, additional scenarios have been put forth such
as multiphase {\it in situ} formation \citep{for97}, major mergers
\citep{zep93} and tidal stripping/capture \citep{cot98}.

Support for the idea that the Milky Way halo formed in a hierarchical
or episodic manner arises from the lack of a significant abundance
gradient throughout the entire halo globular cluster population
\citep{par00}, a notable decrease in the radial component of the
velocity ellipsoid of the stellar halo beyond the solar circle
\citep{som97}, and the detection of a possible spread in cluster and
stellar ages within the outer halo \citep{sar97, ros99, ste99}.  In
addition, some studies have uncovered evidence of kinematic and
chemical sub-clustering within the globular cluster population of the
Andromeda galaxy which may point to accretion remnants \citep{ash93,
sai00}.  However, it is possible that some degree of fragmentation
might arise in an infall scenario due to thermal and gravitational
instabilities within the collapsing proto-galactic gas cloud.  A
better understanding of the fragment properties including chronology,
chemistry and, in particular, kinematics would help us to identify the
origin of sub-clustering and to differentiate between the various
formation models.

It has been noted often that globular clusters (GCs) provide us with
ideal probes of galaxy structure and formation mechanisms.  As the
oldest galactic stellar systems, globular clusters hold the key to
uncovering the formation record of their host environments.
Correlations between chemical content (which points to chronology)
and the kinematics of globular clusters allow us to investigate the
early formative stages of a galaxy and, to some extent, probe its
ongoing evolution.

The M31 globular cluster system provides an obvious and desirable
target of observation: ({i}) its proximity makes it the most
accessible large system outside of the Milky Way; ({ii}) with
$\gtsim435$ confirmed candidate members \citep{bar00}, the M31 GCS is
sufficiently populous that it provides a statistically significant
sample size; ({iii}) unlike the Milky Way GCS, the M31 population can
be observed over the full extent of the galaxy with less severe
effects of line-of-sight contamination; ({iv}) all of the M31 clusters
lie at essentially the same distance, and thus are not subject to such
large individual distance uncertainties as are the Milky Way
globulars.  However, the proximity of M31 introduces certain issues of
contamination within the cluster sample, such as a difficulty in
distinguishing between GC candidates and star-forming regions in many
photometric surveys.
Furthermore, faint globular clusters can easily be missed against
the bright galaxy background.

Previous studies of the Milky Way and M31 globular cluster systems
have revealed that these two populations exhibit some remarkable
similarities.  \citet{zin85} found that the Galactic GCS demonstrated
clear evidence for bimodality in its metallicity distribution, and
proposed that it could be considered as two distinct subsystems: a
flattened, metal-rich, rapidly rotating, kinematically cold disk
system, and a spherical, metal-poor halo with lower net rotation and
higher velocity dispersion.  Signs of bimodality have been found in
the M31 metallicity distribution \citep{ash93} and in
metallicity-sensitive color distributions \citep{bar00}.  Separating
the M31 clusters into metallicity sub-systems along the same lines as
the Milky Way GCS, \citet{huc91} found that the metal-rich clusters
within $R\sim7\arcmin$ form a rapidly rotating ($100-200$~\kms)
disk-like system, whereas the metal-poor clusters within this radius
exhibit no net rotation.  Beyond this radius, they were unable to
distinguish the metal-rich from metal-poor populations on the basis of
kinematics.  Later studies have provided additional evidence
supporting the view that M31 has a rapidly rotating system of
metal-rich globular clusters \citep{ash93, bar00}.

\citet{arm89} confirmed the presence of a modest metallicity
gradient within the system of disk clusters in the Milky Way.  In the
first extensive spectroscopic survey of the M31 GCS, \citet{vdb69}
found no significant metallicity gradient with
position but did report a wide range of abundances at any given
radius.  More recently, Huchra et al.\ (1982, 1991) and \citet{sha88}
have shown that there is evidence for a weak but measurable
metallicity gradient with projected radius in the M31 cluster system.  A
similar finding was reported by \citet{bar00} based on their large
sample of metal-sensitive colors.  \citet{huc91} found a mean M31 GC
metallicity of ${\rm [Fe/H]} = -1.2$, slightly higher than that for
the Milky Way.

There remains a lack of persuasive evidence that would allow us to
distinguish between halo formation models --- or indeed to suggest a
clear hybrid of (or alternative to) the classical formation scenarios.
It is clear that an extensive database of high-precision data for the
kinematic and chemical tracers of a large number of host galaxies is
required in order to help solve this currently intractable problem of
galaxy formation.  Only then can the samples be examined for clear
evidence of kinematic anomalies, metallicity gradients and
sub-structure which, if present, may point to the formation and
enrichment processes which have occurred in the galaxy.  With this
objective, we present spectroscopic results for $\gtsim200$ M31
cluster candidates, adding a large number of new observations to the
database.

This paper will take the following form: Section~\ref{sec:targets}
contains a description of the target selection and coordinate
determinations. In Section~\ref{sec:data}, we provide a summary of the
data acquisition and reduction procedure along with a description of
the velocity and metallicity determinations.  We incorporate these new
results into the broader sample of available M31 cluster data to
provide an analysis of the abundance distribution of the globular
cluster system of our nearest large spiral galaxy neighbor in
Section~\ref{sec:metallicities}.  In Section~\ref{sec:kinematics}, we
examine the kinematics of the cluster system and provide a mass
estimate for the galaxy. A summary and discussion of the results of
this study are presented in Section~\ref{sec:conclusion}.


\section{Target Selection and Coordinates}
\label{sec:targets}

The list of targets was selected from the catalogue of \citet{bat87},
with some additional central targets from unpublished CCD photometry
by J.~Huchra.  Cluster positions accurate to $0\farcs2$ were
determined from Automatic Plate Measuring (APM) scans of deep
photographic plates of a $6^\circ\times6^\circ$ area around M31.  The
accuracy of the APM cluster coordinates represent a marked improvement
over existing M31 GCS positions, which have typically been measured to
the nearest arcsecond.

Galactocentric coordinates for the GCs were computed relative to
an adopted M31 central position of $\alpha_0=00^h40^m00.1^s$,
$\delta_0=+40^\circ59\arcmin43\arcsec$ (B1950):
\begin{eqnarray}
X &=& C_1\,\sin{(PA)} + C_2\,\cos{(PA)} \\
Y &=& -C_1\,\cos{(PA)} + C_2\,\sin{(PA)}, \nonumber
\end{eqnarray}
where $C_1=[\sin{(\alpha - \alpha_0)} \cos{\delta}]$ and
$C_2=[\sin{\delta}\cos{\delta_0} - \cos{(\alpha - \alpha_0)}
\cos{\delta}\sin{\delta_0}]$.  The $X$ coordinate represents the
position along the major axis of the galaxy increasing towards the
north-east along the major axis.  The $Y$ coordinate is the distance
from the major axis above and below the galactic plane, with positive
values towards the north-west.  The position angle (PA) of the major
axis was taken to be $38^\circ$ \citep{ken89}.

Numerous groups have contributed to the identification of the globular
cluster members of M31 \citep{hub32, vet62, sar77, cra85, bat80,
bat87, bat93}.  The ultimate result of this combined effort is that
most of the clusters are known by more than one name, a fact which has
the potential to cause confusion.  The letters preceding the
identification number in the M31 cluster target names given hereafter
indicate the relevant catalogue reference as provided in
Table~\ref{tab:targref}.


\section{The WYFFOS Data}
\label{sec:data}

The data for this study were obtained at the William Herschel 4.2~m
telescope\footnote{The William Herschel Telescope is operated on the
island of La Palma by the Isaac Newton Group in the Spanish
Observatorio del Roque de los Muchachos of the Instituto de
Astrofisica de Canarias.} (WHT) in La Palma, Canary Islands, during the
nights of November 3 to 6, 1996.  The WYFFOS Wide Field Fibre Optic
Spectrograph \citep{bri98} was used with two gratings to provide a
total spectral coverage of $\sim 3700 - 5600$~\AA.  The H2400B
2400-line grating yielded a dispersion of 0.8~\AA/pixel and a spectral
resolution of 2.5~\AA\ over the range $3700-4500$~\AA.  This
wavelength range was selected to investigate the CN feature at
3883~\AA, the H \& K lines of calcium, H$\delta$, the CH G-band and
the 4000~\AA\ continuum break. The R1200R 1200-line grating
observations had a dispersion of 1.5~\AA/pixel and a resolution of
5.1~\AA\ over the spectral range $4400-5600$~\AA\ to add absorption
features such as H$\beta$, the Mgb triplet and two iron lines near
5300~\AA.

\begin{figure*}
\hspace*{\fill}\psfig{file=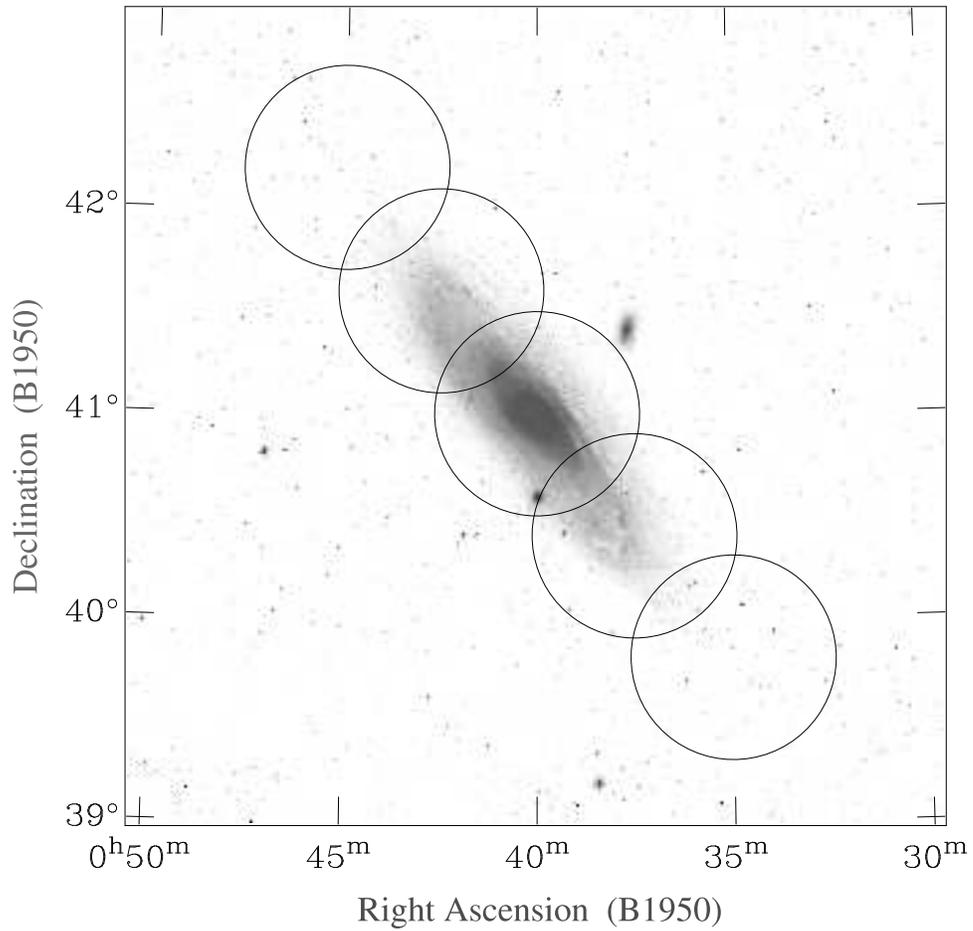,width=5in}
\hspace*{\fill}\caption{Five $1^\circ$ WYFFOS fields shown
superimposed over an optical image of the galaxy from the Digitized
Sky Survey.  Two central fibre configurations were used to yield
spectral observations for a total of 288 cluster candidates.  North is
to the top, east is towards the left.}
\label{fig:wht_fields}
\end{figure*}

Six different fibre configurations were selected to target a total of
288 globular cluster candidates over the full extent of the major axis
of M31.  An overlay of the WYFFOS target fields is shown superimposed
on a Digitized Sky Survey image of the galaxy in
Figure~\ref{fig:wht_fields}.  A record of the observations is provided
in Table~\ref{tab:obslog}, which includes the coordinates of the field
centers, observation dates and exposure times.  Regrettably, minor
axis fields were not observed during this run due in part to a loss of
time from poor weather.

Spectra of twilight flats, argon arc lamps, and velocity template
stars were also obtained during the same nights of observation as the
targets.  The data reduction --- including fibre throughput
corrections, spectral extraction, sky subtraction and wavelength
calibration --- was accomplished using the {\tt wyf\_red} WYFFOS
Multi-fiber IRAF\footnote{Image Reduction and Analysis Facility,
distributed by the National Optical Astronomical Observatories, which
is operated by AURA under contract with the NSF.}  package written by
Jim Lewis.  Ensemble median RMS values for the wavelength calibration
were typically $0.02 - 0.08$~\AA\ for the H2400B grating images
and $0.08 - 0.13$~\AA\ for the R1200R grating.

\begin{figure*}
\hspace*{\fill}\psfig{file=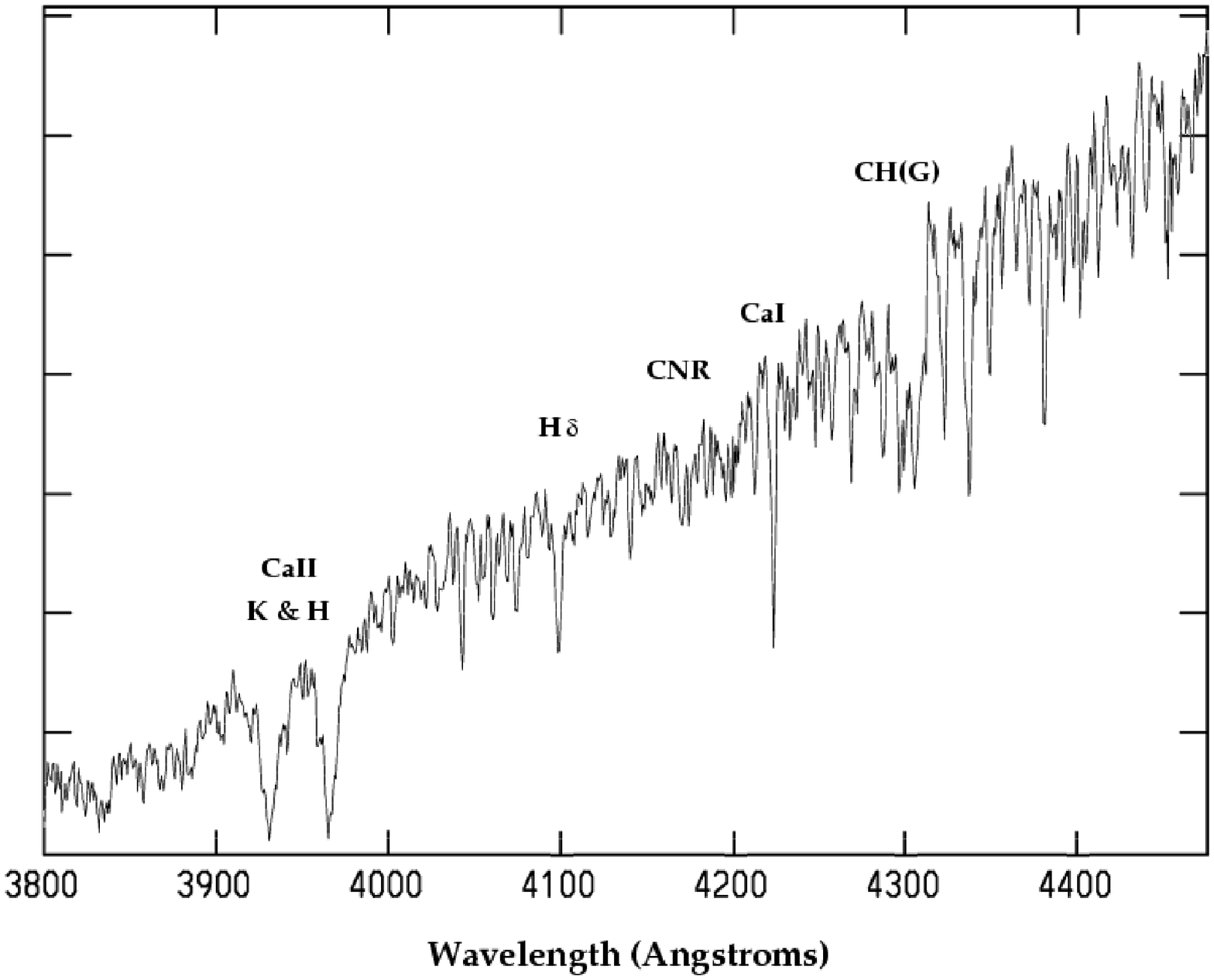,width=4in}
\hspace*{\fill}\caption{A sample reduced spectrum for M31 globular
cluster B19, obtained using the WYFFOS system and the H2400B grating.
The ordinate axis is linear in arbitrary units (zero counts is
at the bottom).}
\label{fig:b19b}
\end{figure*}

\begin{figure*}
\hspace*{\fill}\psfig{file=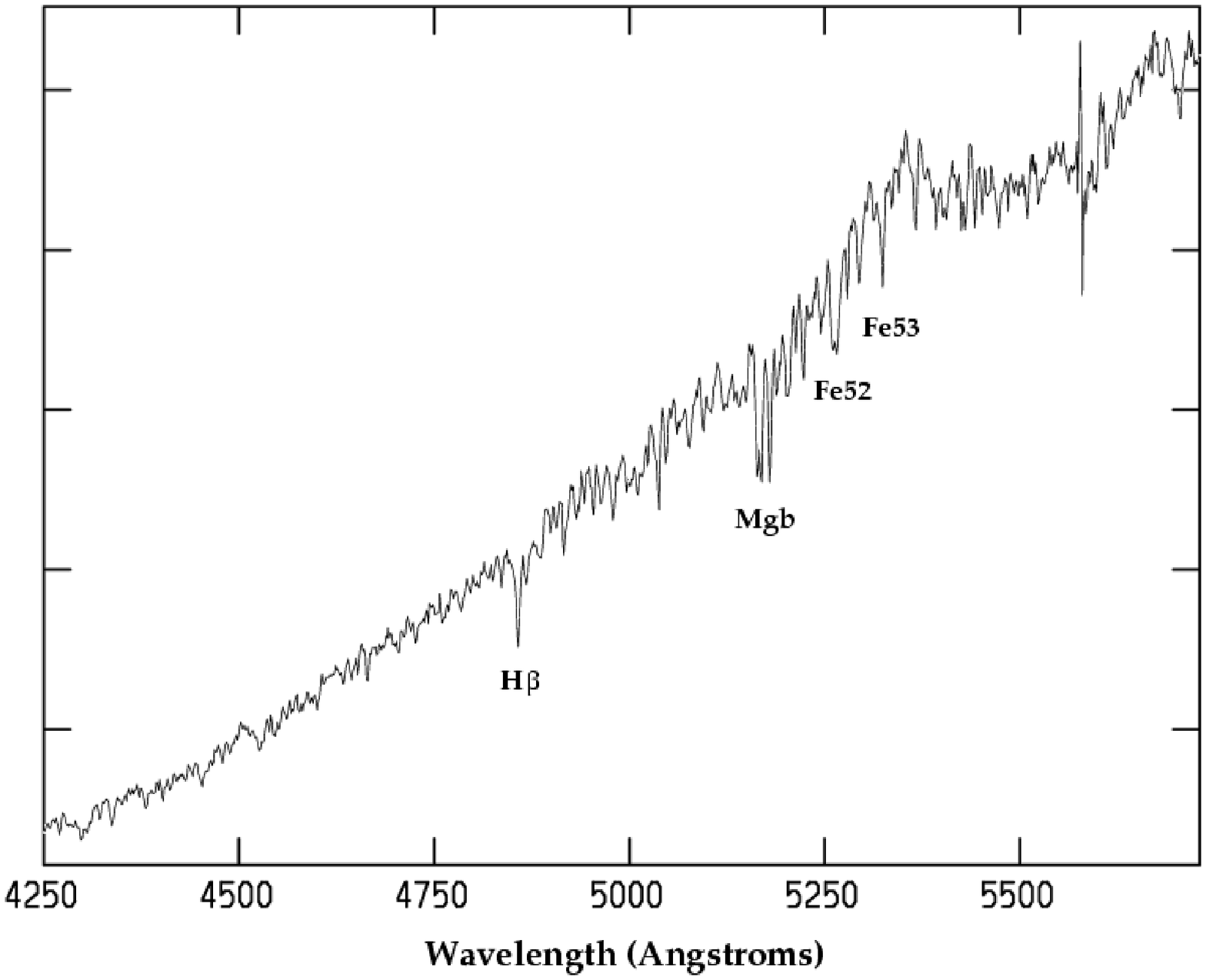,width=4in}
\hspace*{\fill}\caption{A sample reduced spectrum for M31 globular
cluster B19 obtained using the WYFFOS system and the R1200R grating.
The ordinate axis is linear in arbitrary units (zero counts is
at the bottom).}
\label{fig:b19r}
\end{figure*}

Most of the globular clusters in the M31 fields are superimposed on a
relatively bright, non-uniform background of galaxy light.  For the
central WYFFOS fibre configurations, roughly 20 blank-sky fibres were
arranged in a regular pattern across the $1^\circ$ field to
characterize the background intensity.  The outer fields contained
more dedicated sky fibres thanks to a lower density of targets in the
area.  Spectra obtained through the dedicated background sky fibres
were averaged to form a composite sky spectrum which was then
subtracted from each target.  Attempting to match one or a few nearby
sky fibres to each target spectrum (i.e. to associate the local sky
level with each cluster) did not prove advantageous, since the
resulting losses in signal-to-noise largely outweighed any gains in
compensating for background variability even within the central
fields.  Examples of reduced spectra obtained using both gratings are
shown in Figures~\ref{fig:b19b} and \ref{fig:b19r} with the absorption
features of interest labelled.


\subsection{Radial Velocities}

Radial velocities for the M31 globular clusters were calculated by
performing standard Fourier cross-correlations of each target spectrum
with the high signal-to-noise spectra of template stars of known
velocity \citep{ton79}.  Two spectra were acquired for each target,
obtained using each of the R1200R and H2400B gratings.  Two template
stars, HD12029 (K2 giant) and HD23169 (G2 dwarf), were also observed
during the run to yield four cross-correlation combinations for the
derivation of velocities for each cluster target.  We required that
each target yield a minimum of two satisfactory cross-correlations
with peak values $CC\geq0.3$ to qualify as a successful result (a
result of $CC=1$ would indicate a perfect correlation).  A mean of the
cross-correlation results, weighted by the associated Tonry \& Davis
R-value and corrected for solar motion, was used to calculate
heliocentric radial velocities for a total of 202 M31 clusters
candidates in the sample.  The remainder represent:
\begin{enumerate}
\item Confirmed non-clusters (foreground stars, background galaxies,
HII regions, open clusters etc.) identified by spectral features
grossly atypical of globular clusters or radial velocities that were
significantly beyond reasonable limits for the M31 GC population
($|v_{\rm obs} - v_{\rm sys}| > 3\sigma_{v}$ where
$\sigma_{v}\sim150$~\kms).  The status of these objects is noted in the final
column of Table~\ref{tab:WHT}.
\item Target spectra with insufficient signal-to-noise to generate
reliable velocities.
\item Seven objects with inaccurate coordinates due to
misidentifications or saturation on the APM scans.
\end{enumerate}

\begin{figure*}
\hspace*{\fill}\psfig{file=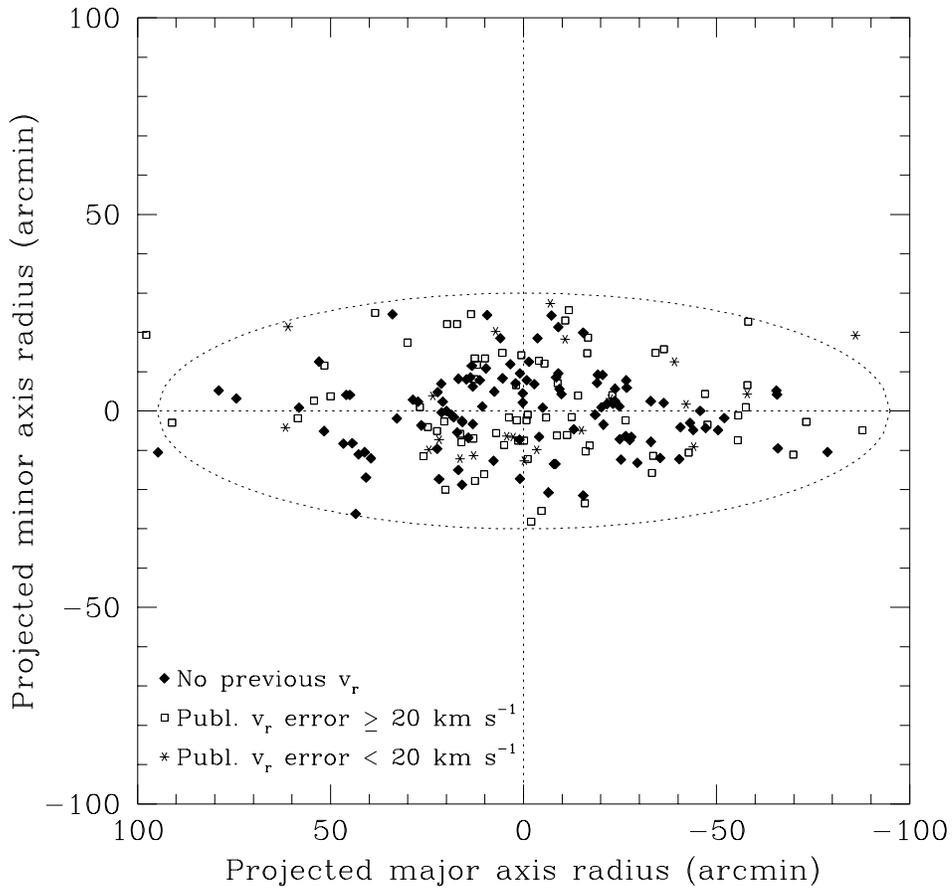,width=5in}
\hspace*{\fill}\caption{The positions of the 202 M31 GCs for which
radial velocities were obtained from the WYFFOS spectra.  Over half of
this sample had no previously published spectroscopic data.  Of those
that had, our velocities (with adopted uncertainties of $\pm 12$~\kms)
represent significant improvements over most of the pre-existing
spectroscopic data.}
\label{fig:whtvelpos}
\end{figure*}

The radial velocities obtained from the WYFFOS spectra are provided in
Table~\ref{tab:WHT}.  Over half (109) of the M31 GCs for which
velocities were obtained had no previously published spectroscopic
data.  The majority of the remaining clusters (72) had published
velocities with large uncertainties of $20-80$~\kms, and the remainder
(21) had modest to high-precision velocities with errors
$<20$~\kms. The positions of the WYFFOS targets in each of these
categories are shown schematically in Figure~\ref{fig:whtvelpos}.

Robust fits revealed no significant differences between the velocities
determined using either template object with a given grating.  A
comparison of the velocities calculated for a given target from the
two gratings showed only slightly larger differences; the derived
velocities found using the R1200R grating data, for example, were at
worst $\sim 6$~\kms\ larger than those determined from the spectra
obtained using the H2400B grating.  An uncertainty of $\pm12$~\kms\ is
adopted here for the M31 clusters in our sample based on the RMS of
the velocity deviations for 28 targets observed in multiple fields.
On average, this represents an overestimate of the true uncertainty
since it incorporates certain internal systematic errors as well as
statistical errors.

\begin{figure*}
\hspace*{\fill}\psfig{file=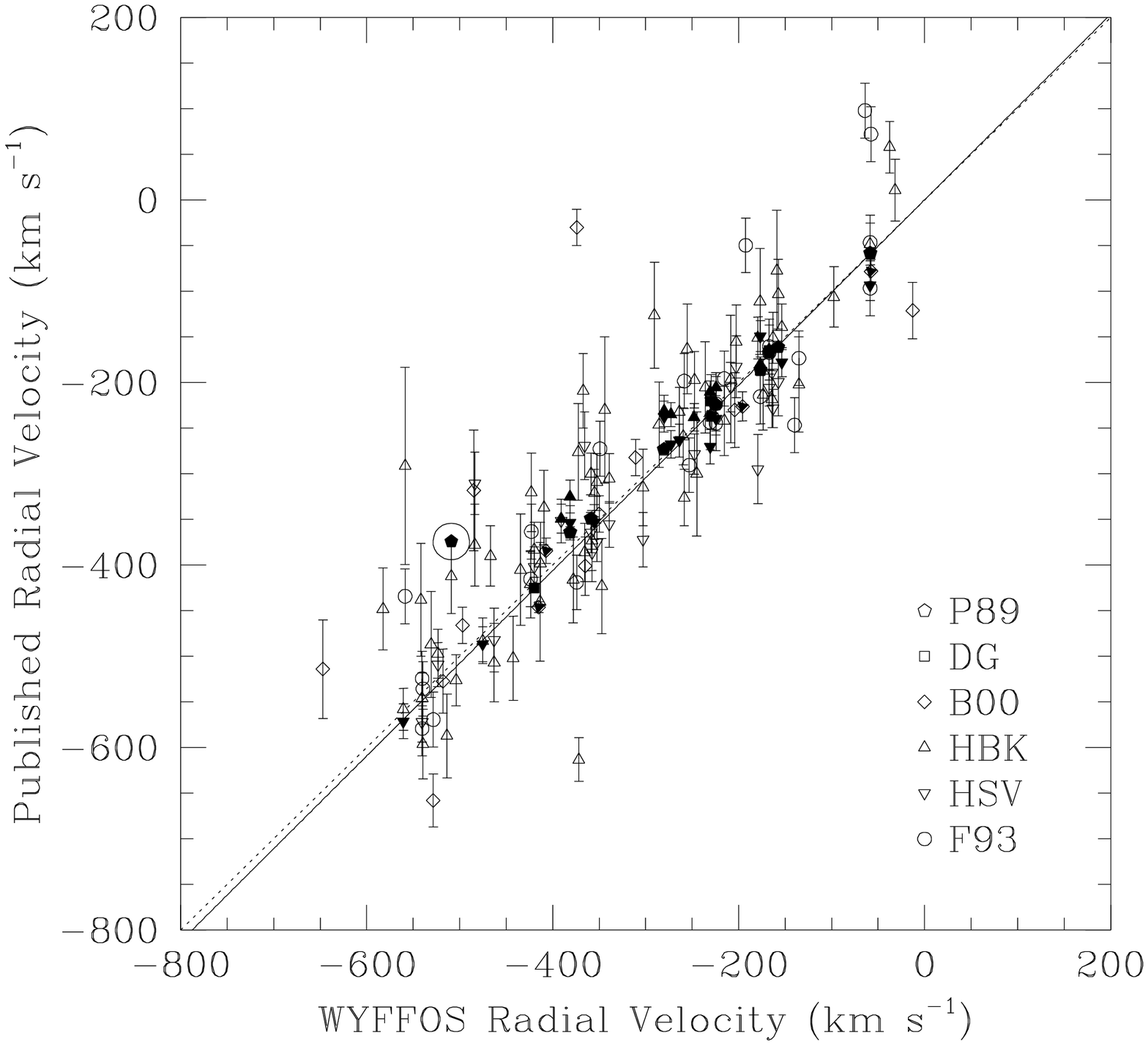,width=5in}
\hspace*{\fill}\caption{A comparison between radial velocity results
obtained from this study and those in other publications: P89 =
\citet{pet89}; DG = \citet{dub97}; B00 = \citet{bar00}; HBK =
\citet{huc91}; HSV = \citet{huc82}; F93 = \citet{fed93}.  Filled
points represent high-precision velocity determinations with quoted
errors less than $20$~\kms.  The unity relation is represented by the
dotted line, and the results of a linear fit with a $1/\sigma^2$
weighting is shown by the solid line.  The circled (filled) point is
M31 globular cluster B29 (see text for details).}
\label{fig:publ_vels}  
\end{figure*}

Figure~\ref{fig:publ_vels} shows a comparison of those radial
velocities in the present study that overlap with other samples from
the literature.  A linear fit, weighted by the inverse square of the
published velocity uncertainties, yields $v_{\rm publ} = 0.43 +
1.02\,v_{\rm obs}$~\kms.  Overall, there is good agreement between the
WYFFOS velocities derived herein and those of other studies within the
adopted errors.  Encouragingly, the RMS residuals of independent fits
to the separate velocity samples were particularly low for the
clusters which overlap with the high-precision echelle studies of
Peterson (1989: $v_{\rm P89} = -5.8 + 0.96\,v_{\rm obs}$~\kms\ with
${\rm RMS}=5.7$~\kms\ for 8 objects) and Dubath \& Grillmair (1997:
$v_{\rm DG} = 0.58 + 1.02\,v_{\rm obs}$~\kms\ with ${\rm
RMS}=8.2$~\kms\ for 5 objects).  Note that the only high-precision
result that deviates significantly from our fit is for B29, a cluster
for which \citet{pet89} notes her velocity may be incorrect.  This GC
was not used in the calculation of the weighted fit or RMS, and it is
circled in Figure~\ref{fig:publ_vels}.

There are two other points in Figure~\ref{fig:publ_vels} which deviate
significantly from the fit, and thus deserve mention.  The open
diamond above the best fit line represents the M31 globular cluster
B301-S22: \citet{bar00} obtained $v_r=-30\pm 20$~\kms, in poor
agreement with the value of $-374\pm 12$~\kms\ measured in the current
study.  The Barmby et al.\ coordinates match well with the APM
coordinates for this object as provided in Table~\ref{tab:WHT}, so the
velocity deviation does not appear to be due to a target
misidentification in either study.  This cluster was also measured by
\citet{fed93}, who obtained a velocity of $-419 \pm 30$~\kms, in good
agreement with the WYFFOS result.  The Barmby et al.\ sample also had
the largest RMS residuals in the comparison fits: ${\rm
RMS}=110$~\kms\ for 16 objects, not including B301-S22.  The reasons
for this large discrepancy are not known.  The open upright triangle
which deviates from the fit in Figure~\ref{fig:publ_vels} represents
B109-S170, for which \citet{huc91} measured a velocity of $-613\pm
24$~\kms\ (compared to $-372\pm 12$~\kms\ from the WYFFOS results).
Again, the coordinates for this object match well between both studies
and the deviation does not appear to be due to a target misidentification.


\subsection{Absorption and Reddening Corrections}

Each spectrum was corrected for the effects of atmospheric absorption
and for reddening in both the Milky Way and in M31 itself.  Color
excess values upon which the reddening calculations are based were
determined using the slope ($S$) of the continuum between
$\sim4000-5000$~\AA\ as tabulated in \citet{cra85}, where available:
\begin{equation}
\label{eq:crampton4}
E_{B-V}= -0.066\,S + 1.17\,(B-V) - 0.32.
\end{equation}
The slope parameter has been shown to be a good estimator of
reddening, and is not very sensitive to cluster metallicity
\citep{els88}.

Targets with published $(B-V)$ colors but with no slope parameters or
intrinsic color estimates were de-reddened to a typical color of
$(B-V)_0=0.76$ for M31 globular clusters at large galactocentric radii
\citep{cra85}.  Clusters with no $S$ values (or colors) and those for
which the calculated excess values were lower than a minimum
foreground color excess of $E_{B-V} = 0.10$ (due to errors in the
slope parameter or in the photometry) were assigned this minimum value
\citep{fro80,els88}.


\subsection{Abundance Determinations}

To rank the M31 clusters in the WYFFOS sample by metallicity, we have
measured the strengths of various absorption features in their
integrated spectra.  Line indices were calculated to measure the
signal in a wavelength-delineated region centered on each spectral
feature relative to the signal in the blue and red continuum zones
flanking the feature.  In light of the fact that no flux standard star
spectra were obtained during the WYFFOS observing run, it was not
possible to convert photon counts directly into flux for these
targets.  Our motivation, however, was to obtain a metallicity ranking
of the globular clusters in M31; relative flux calibration was not
essential as its absence will not significantly affect narrow features
and mean count ratios were used instead.  Our line and continuum
bandpasses were defined as shown in Table~\ref{tab:LIs} \citep{bur84,
bro86, bro90}.

Each target spectrum was shifted to zero radial velocity to define
consistent line index wavelengths in the rest frame.  The velocity
dispersion of the M31 globular cluster system is roughly $150$~\kms, a
spread which translates into a nominal range of $4\sigma_v\gtrsim
600$~\kms\ in velocity.  This would correspond to a wavelength shift of
$\sim 9$\AA\ at $\lambda=4500$~\AA, a significant fraction of the
width of most of the bandpasses selected for measurement.  Thus,
metallicity determinations were not pursued for target spectra which
had yielded no velocity cross-correlations.

\noindent The generalized feature index is defined as
\begin{equation}
I = -2.5\log{\left[\frac{2\bar\protect{\cal F}_I} {\bar\protect{\cal
F}_{C1}+\bar\protect{\cal F}_{C2}}\right]}
\end{equation}
where $\bar{\cal F}_{I}$ , $\bar{\cal F}_{C1}$ and $\bar{\cal F}_{C2}$
represent the mean count levels in the feature bandpass and associated
continuum bandpasses $C1$ and $C2$ as given by $\bar{\protect{\cal F}}
= (\lambda_2-\lambda_1)^{-1}\int_{\lambda_1}^{\lambda_2}\protect{\cal
F}\,d\lambda$ \citep{bro86, bro90}.  The continuum break feature
$\Delta = 2.5\log\left[{\bar{\cal F}_{C2}}/{\bar{\cal F}_{C1}}\right]$
is a measure of the discontinuity in the continuum due to Fraunhofer
line-blanketing shortward of 4000~\AA.  This feature is somewhat more
vulnerable to errors in the reddening and absorption corrections, as
it spans a broad range in wavelength and is asymmetric.

\begin{figure*}
\hspace*{\fill}\psfig{file=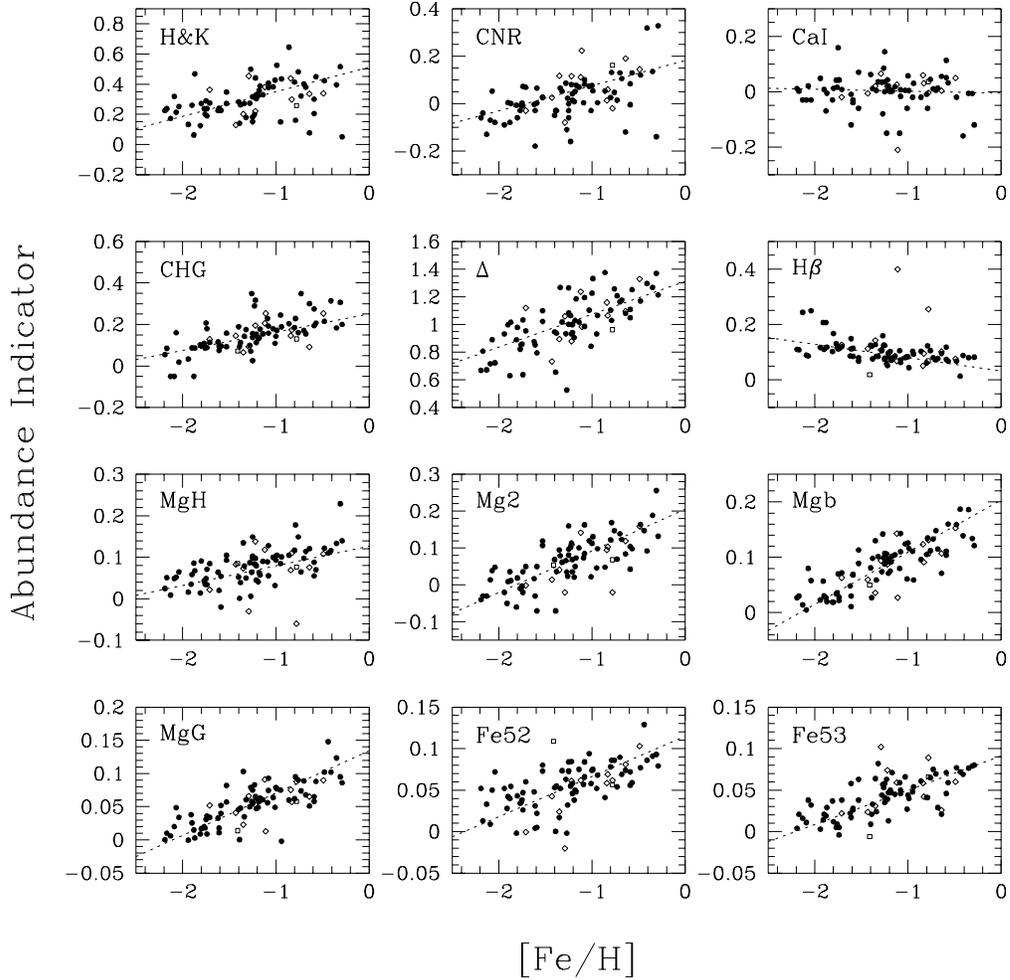,width=5.5in}
\hspace*{\fill}\caption{The relationship between published line [Fe/H]
values and the line indices in this study.  Metallicity sources are as
follows: $\bullet$ Huchra et al 1991, $\diamond$ Barmby et al 2000 and
$\Box$ B\`onoli et al 1987.  The dotted lines represent linear fits to
the data, weighted by $1/\sigma_{\rm [Fe/H]}^2$.}
\label{fig:FeHfits}
\end{figure*}

Figure~\ref{fig:FeHfits} shows a comparison of our line indices with
published M31 globular cluster [Fe/H] values from various sources
\citep{bon87, huc91, bar00}.  The RMS differences between the line
indices calculated from spectra observed in more than one field
through different fibres were used to estimate the internal errors for
each feature, and these are shown in the second column of
Table~\ref{tab:FeHfits}.  The results of linear fits in the form
$LI=a+b{\rm [Fe/H]}$ weighted by {$1/\sigma_{\rm [Fe/H]}^2$} are
provided in columns 3 and 4 of Table~\ref{tab:FeHfits}.  Note that
only spectroscopic metallicities for overlapping targets were
considered in the calibration and hence the \citet{bon87}
near-infrared photometric metallicities were not used in the fits.  In
column 5 of Table~\ref{tab:FeHfits}, we provide the RMS residuals of
the fit and in column 6 we list the strength of the $r$ coefficient, a
measure of the linear correlation between the line indices and
published metallicities.

To best represent the metal abundances of the WYFFOS targets observed,
a line index was selected to contribute to the calibration if it met
the following four criteria:

\begin{enumerate}

\item The uncertainties associated with the line index did not exceed
$\sim20\%$ of the full range of observed values.  The Fe52 index did
not satisfy this condition and was thus rejected from the calibration.

\item The linear correlation coefficient $r$ exceeded a minimum
level of $r=0.65$.

\item Numerical experimentation demonstrated that the feature strength
was stable in the presence of significant changes in the adopted value
of cluster reddening (e.g.~in comparison with any adopted reddenings
which differed from unpublished M31 values from \citet{bar00}).  The
continuum break was found to be susceptible to reddening uncertainties
and was thus rejected.

\item Only one line index that characterizes a given spectral feature
was incorporated in the estimate.  In the case of the magnesium
triplet, the indices MgG, Mg2 and Mgb all provided an acceptable
measure of essentially the same absorption feature.  Of these, the Mgb
index demonstrated the highest correlation coefficient and a
relatively low RMS uncertainty and was therefore selected for use
over the others.

\end{enumerate}

\begin{figure*}
\hspace*{\fill}\psfig{file=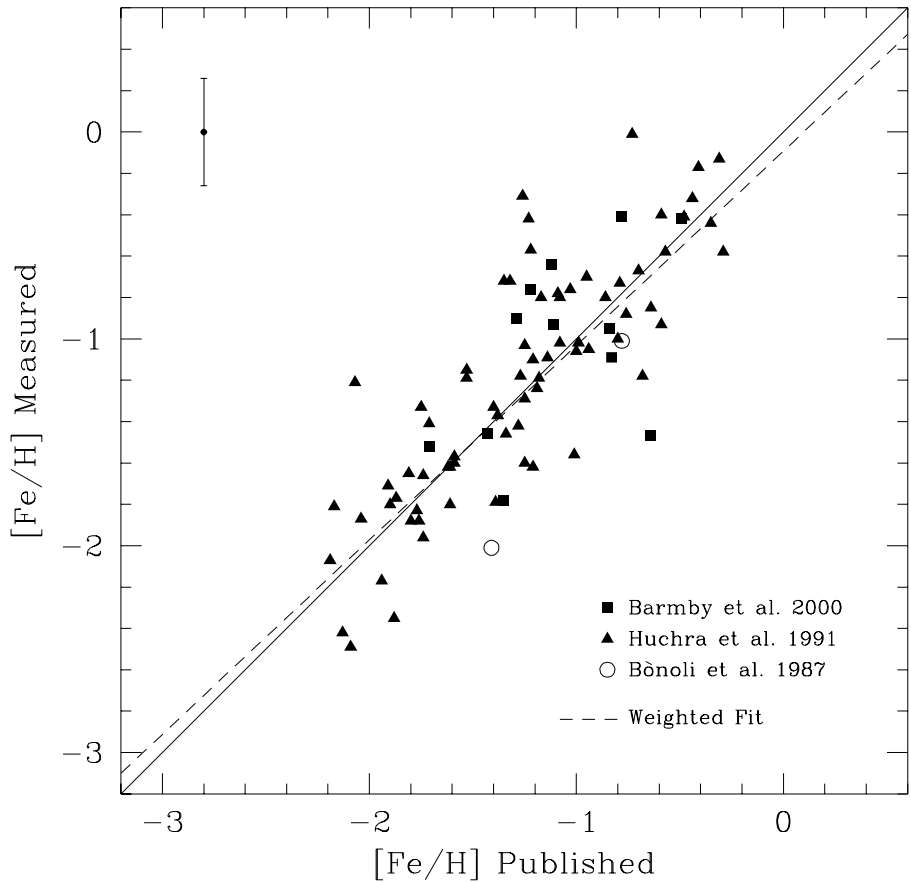,width=5in}
\hspace*{\fill}\caption{The relationship between published line [Fe/H]
values and the calibrated WYFFOS metallicities.  The error bar in the
upper left represents the median formal error quoted on the WYFFOS
values.  A linear fit, weighted by the inverse-square of the
individual uncertainties, is shown by the dashed line for comparison
to the unity relation (solid line).  The slope of the fit is
$0.94\pm0.02$ and the RMS of the fit residuals is 0.24 dex.}
\label{fig:FeHcompare}
\end{figure*}

Line indices which met the above criteria include the CH (G-band)
feature, Mgb to characterize the strength of the magnesium $b$
triplet, and the Fe53 iron line.  Final cluster metallicities have
been determined from an unweighted mean of the [Fe/H] values
calculated from the CH(G), Mgb and Fe53 line strengths for 202 of the
WYFFOS targets.  These metallicities and their associated errors are
provided in columns 7 and 8 of Table~\ref{tab:WHT}.  A plot of the
correlation between our [Fe/H] values and those in other samples from
the literature is presented in Figure~\ref{fig:FeHcompare}.  The
best-fit of the data in Figure~\ref{fig:FeHcompare} has a slope of
$0.94\pm0.02$ and the RMS of the fit residuals is $0.24$.  The median
value of the formal errors on our WYFFOS metallicities is $\pm
0.26$~dex.


\subsection{The Best Current Sample}

In the subsequent analysis, our results were merged with the
velocities and [Fe/H] values from other sources in order to produce
the best available sample of spectroscopic data.  Wherever data from
other publications overlapped with our WYFFOS sample {\it and}
demonstrated smaller associated uncertainties, these data superseded
the WYFFOS results; such cases are noted in the final column of
Table~\ref{tab:WHT}.  In this final sample, 191 WYFFOS velocities and
189 metallicities were combined with data from \citet{pet89},
\citet{huc91}, \citet{fed93}, \citet{dub97}, and \citet{bar00} to
yield a total M31 globular cluster spectroscopic database comprising
321 velocities and 301 metallicities.

\citet{bar00} identify the following targets as having spectral
properties which are inconsistent with globular clusters: B55-S116
(stellar), B308 (galaxy), B341-S81 (stellar) and B392-S329 (stellar).
We concur with the identification of B308 as a definite non-cluster
and with that of B392-S329 as a probable non-cluster, and do not carry
these through the remainder of the analysis.  We do not find any clear
abnormalities in the spectra of the other targets and thus cannot at
present reject these as confirmed non-clusters.

\begin{figure*}
\hspace*{\fill}\psfig{file=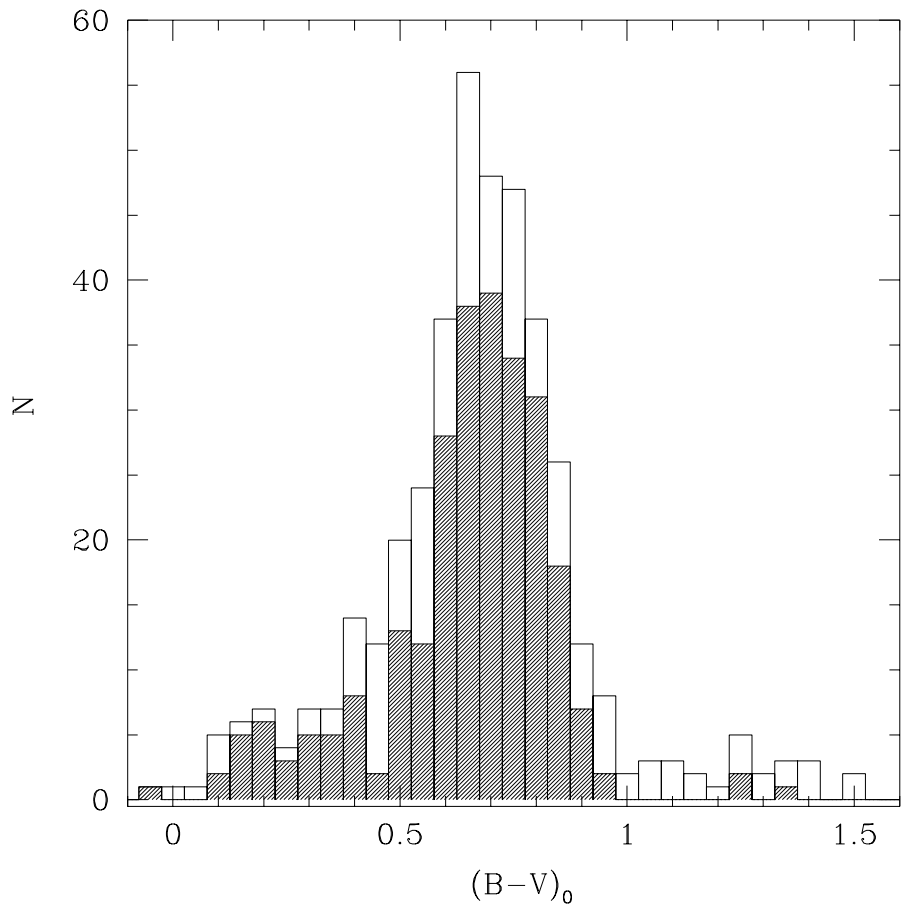,width=5in}
\hspace*{\fill}\caption{The color distribution of the M31 globular
clusters for which we have spectroscopic metallicities (shaded) as
compared to the overall color distribution of the GCS (solid line).
There is no sign of color bias in the metallicity sample as compared
to the overall M31 cluster population with available $(B-V)$ color
information \citep[from][]{bar00}.}
\label{fig:colourdist}
\end{figure*}

In an effort to verify that there are no obvious biases in the
metallicity sample studied here, we examine the color distribution of
our sample and compare it to that of the overall M31 GCS.  The mean
estimated intrinsic color of 412 M31 globular clusters with $(B-V)$
data from \citet{bar00} was $(B-V)_0 = 0.67 \pm 0.04$, while that for
the sample of those with [Fe/H] values used in the metallicity
analysis presented herein (264 clusters with colors) was determined to
be $(B-V)_0 = 0.65 \pm 0.05$.  Based on this and a comparison of the
color histograms for the metallicity sample and the overall GCS
(Figure~\ref{fig:colourdist}), we conclude that there are no
significant population biases with respect to the color of the
metallicity sample presented herein.  A Kolmogorov-Smirnov two-sample
test on the distributions in Figure~\ref{fig:colourdist} confirms what
the eye suggests: the color distribution of the metallicity sample is
drawn from the same parent population as the overall color sample at a
high confidence level (99.5\%).

Clearly, the globular clusters at small galactocentric radii will
suffer from more severe effects of incompleteness due to obscuration
by the disk and bulge of M31.  Depending on the color/metallicity
distribution as a function of position within the galaxy, this
observational effect will conceivably bias the observed metallicity
sample.  We will return to this point in Section~\ref{sec:distn}.


\section{M31 Cluster Metallicities}
\label{sec:metallicities}

\begin{figure*}
\hspace*{\fill}\psfig{file=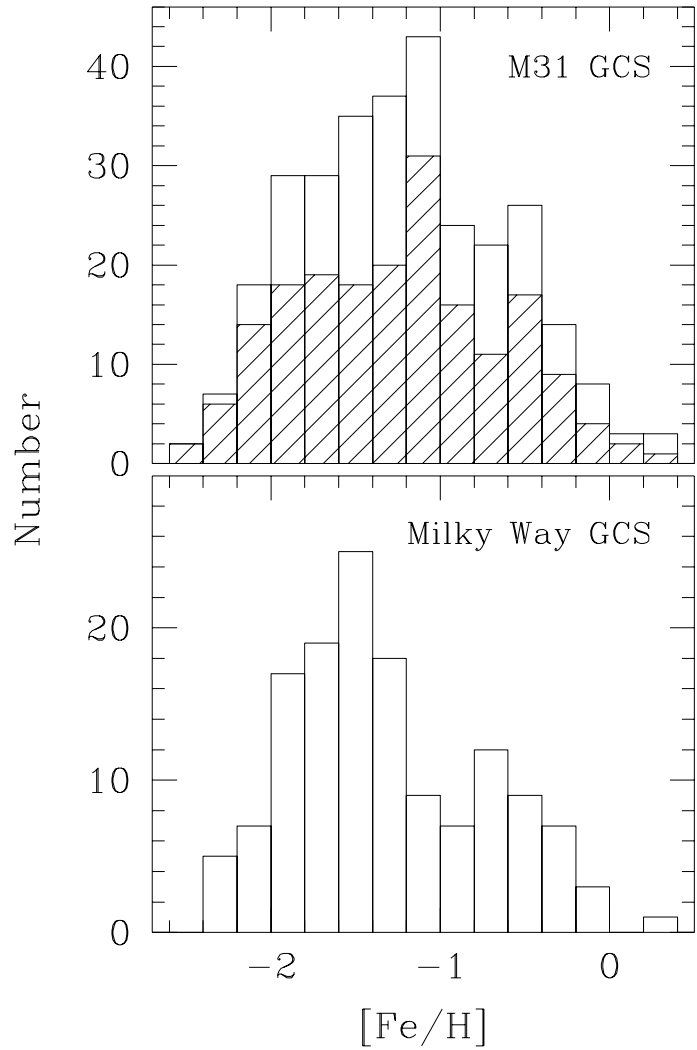,width=6in}
\hspace*{\fill}\caption{The metallicity histogram for the M31 cluster
system (top) and the Milky Way GCS (bottom) for comparison.  The
shaded area in the top plot represents the M31 WYFFOS data.}
\label{fig:fehdist}
\end{figure*}

The metal abundance properties of M31's globular clusters are of
interest as they bespeak the galaxy's formative processes and
enrichment history.  The metallicity distribution for the M31 GCS is
presented in Figure~\ref{fig:fehdist}, along with a similar plot for
the Milky Way GCS for comparison.  The mean metallicity of the full
complement of 301 M31 globular clusters is ${\rm
[Fe/H]}=-1.21\pm0.03$, comparable to the value of $-1.27\pm 0.05$
obtained for the Milky Way GCS (based on data supplied in
\citet{har96}, revised 1999).  However, a Kolmogorov-Smirnov test
reveals that the Galactic and M31 globular clusters have [Fe/H]
distributions which are drawn from the same parent population at only
the 55\% confidence level.

As is frequently the case with binned data, the appearance of the
histogram can be ambiguous and potentially misleading, and thus a more
robust method of analysis is desirable.  In order to examine the shape
of the metallicity distribution without relying on binning methods, we
turned to the KMM algorithm used by \citet{ash94}.  KMM mixture modeling
operates under the assumption that the data is independently drawn
from a parent population which consists of a mixture of $N$ Gaussian
distributions.  The KMM test has previously been applied to the M31
globular cluster metallicity distribution by \citet{ash93}, albeit
with a significantly smaller data sample.

A bimodal distribution was tested against the assumption of a unimodal
shape (the null hypothesis) for the GCS metallicities.  Since there is
no obvious reason to assume that both [Fe/H] groups in a bimodal fit
would exhibit a common covariance, heteroscedastic fitting was applied
such that the covariances for the two distributions were not
constrained to be identical.  The KMM results revealed that this
2-group fit demonstrated a marginal improvement over a single group
fit, although \citet{ash94} cautioned that the output probability in
the heteroscedastic case is more difficult to interpret and may, in
and of itself, be suspect.  We repeated the KMM test on more than 1500
simulations with bootstrap resampling and found that the bimodal case
was preferred over the unimodal distribution at a median confidence
level of higher than $97\%$.  The bimodal test returned Gaussian fits
characterized by means at ${\rm [Fe/H]}=-1.44$ ($\sigma^2=0.22$) for
the metal-poor distribution and at ${\rm [Fe/H]}=-0.50$
($\sigma^2=0.13$) for the metal-rich clusters.  The {\it a posteriori}
probabilities of group membership returned by KMM assigned 231
clusters to the metal-poor distribution and 70 to the metal-rich
population.

In their analysis, \citet{els88} treated the M31 GCS as comprising
three distinct metallicity sub-populations: metal-poor, intermediate,
and metal-rich.  We employed the KMM test to ascertain whether a
trimodal distribution was also consistent with the measured [Fe/H]
values for the M31 cluster system.  The trimodal KMM results with
bootstrap resampling demonstrated that a 3-group fit was indeed
preferred over a single Gaussian distribution at a relatively high
confidence level (96\%). The output population means obtained for the
best 3-group fit were located at ${\rm [Fe/H]}=-1.59$
($\sigma^2=0.19$), $-1.18$ ($\sigma^2=0.18$) and $-0.47$
($\sigma^2=0.12$).  The numbers of clusters assigned to each group
were $N=$153, 76 and 72, respectively.

In light of the variation in the magnitude of [Fe/H] uncertainties
quoted within the WYFFOS sample and others, it is appropriate to test
the effects of these uncertainties on the inferred shape of the
distribution.  There are 229 GCs with metallicity uncertainties lower
than 0.45~dex, a value roughly equal to the variance of one of the
populations as determined from the bimodal results.  We applied the
KMM test to this higher-precision sample to test the stability of our
earlier conclusions.  The bimodal case was again found to represent a
statistically significant improvement over the unimodal assumption
with peaks similar to those found when we incorporated the complete
sample.  The input assumption of three peaks, however, did not return
adequate fits: the {\it a posteriori} probabilities of group
membership in the trimodal case allocated only two clusters to the
intermediate population.

There is an unavoidable smearing of the metallicity distribution due to
metallicity calibration uncertainties, absorption and reddening of
bulge/disk clusters and other line-of-sight problems, or errors
incurred as a result of combining different datasets.  As a further
test, simulations were run in which we shuffled the high-precision
[Fe/H] sample by a random fraction of their associated uncertainties.
This returned a result of marginal significance (median p-value $\sim
0.1$) for the 2-group case, and an insignificant result (p-value $\sim
0.4$) for the 3-group heteroscedastic trial.  Note that the shapes of
the 2-group Gaussian fits were once again consistent with those of the
original bimodal test using the full sample.

\begin{figure*}
\hspace*{\fill}\psfig{file=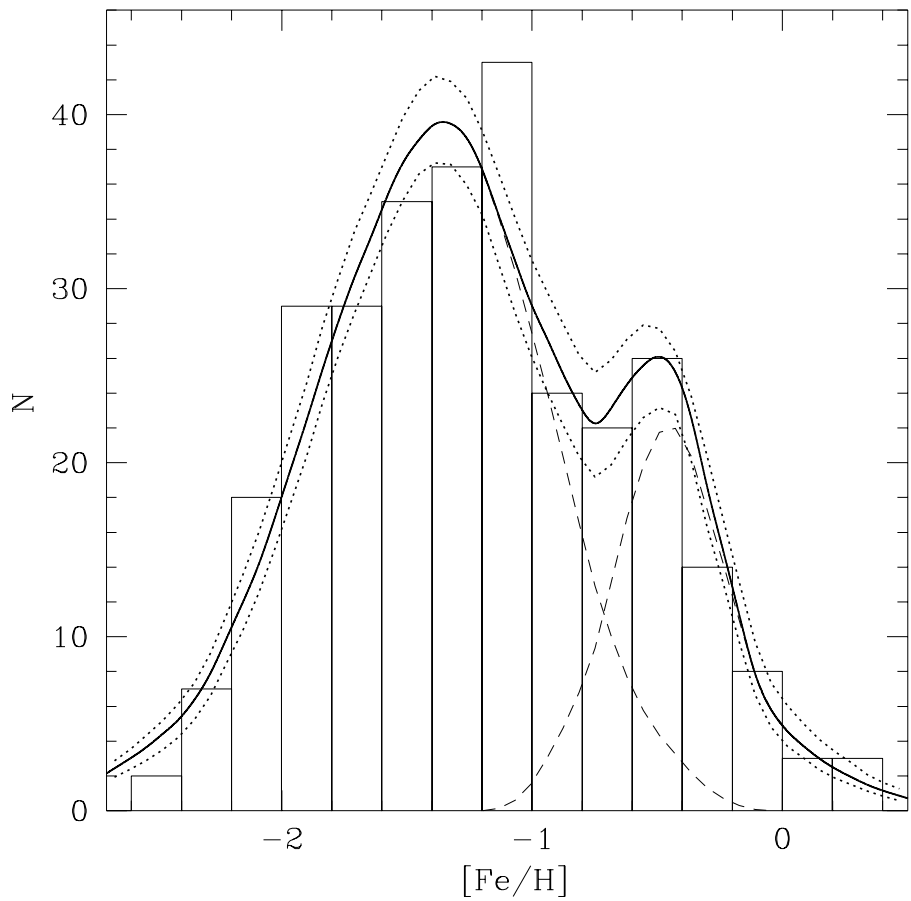,width=5in}
\hspace*{\fill}\caption{Adaptive kernel fits to the M31 metal-poor and
metal-rich GC populations (dashed lines), separated according to the
KMM mixture modeling results.  The sum of the individual fits is shown
(solid line) along with its 90\% confidence interval (dotted
lines). }
\label{fig:bimodal}
\end{figure*}

In light of these KMM results, we tentatively reject the possibility
of a trimodal GC metallicity distribution for the purposes of the
analysis which follows.  While it may yet be reasonable to separate
the cluster system into more than two populations --- say, by culling
out the intermediate targets with group membership probabilities
between 0.4 and 0.6 in the bimodal case --- we elect not to do so
here, as it is not clear that these represent a {\it bona fide}
intermediate population.  Based on the available data, we argue that
the metallicity distribution of the M31 globular cluster system is not
unimodal and better resembles a bimodal distribution similar in
general shape to that of the Milky Way \citep{zin85}.  Therefore,
``metal-poor'' and ``metal-rich'' populations shall refer to those
objects separated into two groups based on the {\it a posteriori}
probabilities returned from the 2-group heteroscedastic KMM test with
the complete available sample.  Each population was independently fit
using an adaptive kernel smoothing procedure\footnote{Adaptive kernel
smoothing code kindly provided by Karl Gebhardt.} and the results are
shown in Figure~\ref{fig:bimodal}.

The positions of our adopted metallicity peaks (${\rm [Fe/H]}=-1.44$
and $-0.50$) differ only slightly from those quoted by \citet{bar00},
who found a bimodal distribution in M31 GC metallicities and
metallicity-sensitive colors with peaks at ${\rm [Fe/H]}=-1.4$ and
$-0.6$.  Our peaks are also similar to the values of ${\rm
[Fe/H]}=-1.5$ and $-0.6$ obtained by \citet{ash93} using the KMM test
on an earlier sample of 144 M31 globular clusters; their study yielded
a ratio of metal-rich to metal-poor globular clusters of 49:95, in
comparison with a ratio of 70:231 found in the current study.


\subsection{Spatial Distribution}
\label{sec:distn}

\begin{figure*}
\hspace*{\fill}\psfig{file=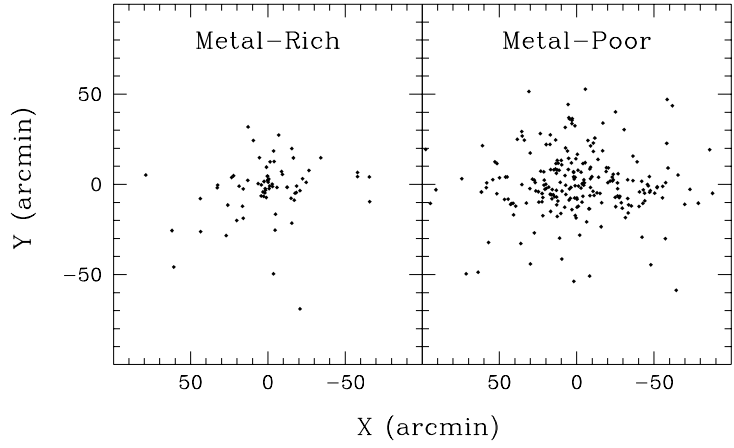,width=4in}
\hspace*{\fill}\caption{The spatial distributions of the metal-rich
and metal-poor cluster populations.  At an M31 distance of 770~kpc,
$5\arcmin$ corresponds to just over 1 kpc.}
\label{fig:spatialdist}
\end{figure*}

The projected spatial distributions of the metal-poor and metal-rich
cluster populations are shown in Figure~\ref{fig:spatialdist}.  The
metal-rich population is clearly more centrally concentrated, an
effect which has been demonstrated previously with a smaller dataset
\citep{huc91}.  In the Milky Way, the metal-rich clusters have been
ascribed as belonging to a thick disk system \citep{zin85, els88,
arm89}.  Later work has suggested that this is better described as a
bulge/bar cluster system \citep{min95, cot99, for01}.  The spatial
concentration and distribution of the metal-rich clusters in M31
appear to be consistent with the latter interpretation.

\begin{figure*}
\hspace*{\fill}\psfig{file=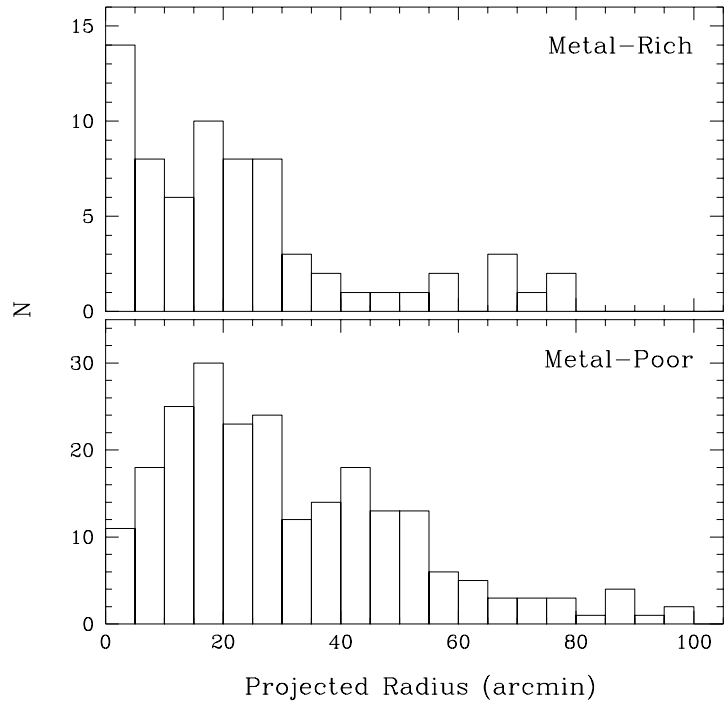,width=4in}
\hspace*{\fill}\caption{The radial distribution of M31 globular
cluster populations with metallicities.}
\label{fig:radprof}
\end{figure*}

There is a notable deficit of metal-poor clusters in the innermost
radial bins shown in Figure~\ref{fig:radprof}.  The already strong
central concentration in the metal-rich population might mask any
analogous deficit in this population.  A similar lack of central
clusters was reported by \citet{els88} within the inner $15\arcmin$ of
the GCS and is at least partially attributable to sampling
restrictions towards the middle of the galaxy, as well as to
observational biases incurred as a result of obscuration by the bulge
and disk.  The cluster sample at large galactocentric radii is bound
to be more complete at faint magnitudes due to the lower degree of
absorption as compared with the inner cluster system.

\begin{figure*}
\hspace*{\fill}\psfig{file=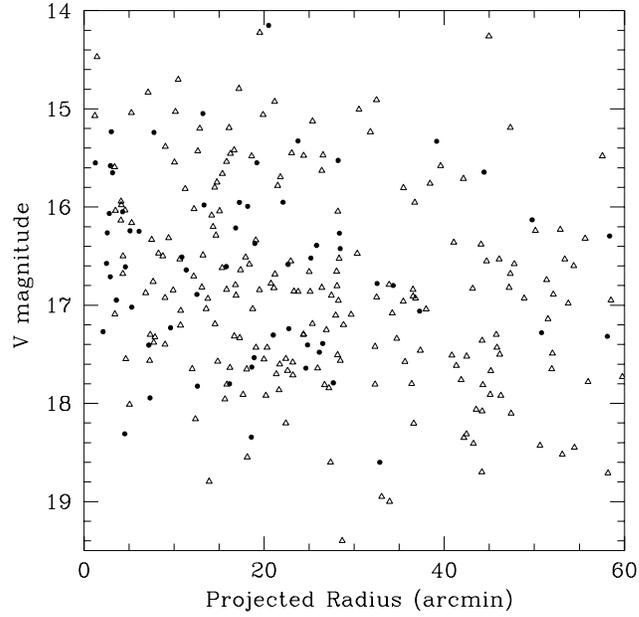,width=3.5in}
\hspace*{\fill}\caption{Apparent V magnitude of the clusters as a
function of galactocentric radius, showing the metal-rich clusters
({\it filled circles}) and the metal-poor clusters
({\it open triangles}).}
\label{fig:vmag_r}  
\end{figure*}

\begin{figure*}
\hspace*{\fill}\psfig{file=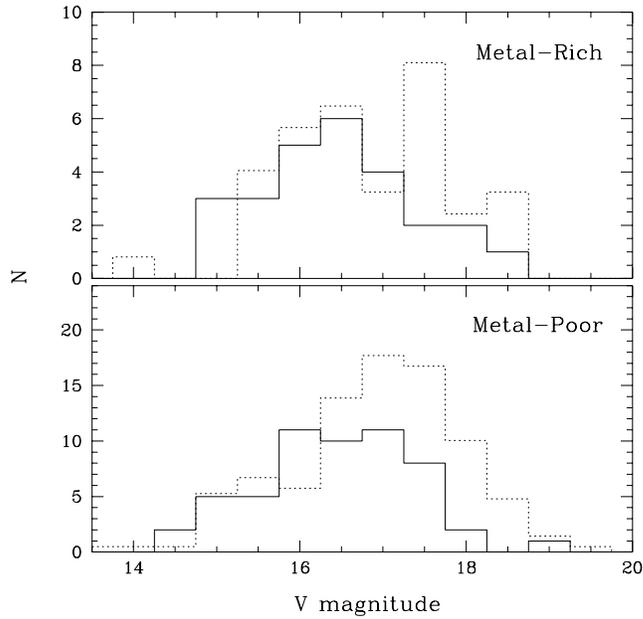,width=3.5in}
\hspace*{\fill}\caption{The observed luminosity functions of the
metal-rich and metal-poor clusters within the inner $15\arcmin$ of the
galaxy are represented by the solid lines.  The outer cluster LFs,
scaled as described in the text, are shown by the dotted lines.}
\label{fig:LF}
\end{figure*}

Examination of Figure~\ref{fig:vmag_r} reveals a modest lack of
clusters fainter than $V\approx17$ within the inner $10\arcmin$ or so
of the GCS.  We attempt to estimate the incompleteness at small radii
in the following manner.  Using available $V$ photometry
\citep{bar00}, we plot the observed luminosity functions (LFs) for the
metal-rich and metal-poor clusters beyond a galactocentric radius of
$15\arcmin$.  These outer LFs are normalized to the same scale as the
distributions within the inner $15\arcmin$ by summing under the
brighter end of the inner and outer histograms ($V<16.75$~mag) and
multiplying by the ratio of these values.  The observed histograms for
the inner metal-rich and metal-poor clusters are shown by the solid
lines in Figure~\ref{fig:LF}, and the scaled outer histograms are
represented by the dotted lines.  Note that this method provides only
a very crude estimate of the number of missing clusters at $V\geq
16.75$ as it presumes, perhaps falsely, that the shape of the globular
cluster LF does not change within the inner region of the galaxy.
Dynamical destruction will have a more severe effect on the inner
clusters and will thus change the overall shape of the luminosity
function.  Evidence for this effect has previously been noted by
\citet{kav97}.

From the scaled outer LFs in Figure~\ref{fig:LF}, we infer that there
are possibly $\sim 30$ missing metal-poor clusters at $V\geq 16.75$,
whereas there may be only $\sim 8$ faint, metal-rich clusters which
are missing within the inner $15\arcmin$ of the galaxy.  Therefore,
sample incompleteness is a reasonable explanation for the dip in the
central bins of the metal-poor radial profile.

Comparing simple counts of the cluster populations within quadrants
centered on the major and minor axes reveals that there is a slight
enhancement in the minor axis fields for the metal-rich population: 40
clusters in the East-West sectors versus 30 in the North-South.  If
one takes into account only the inner $30\arcmin$ of the GCS, this
enhancement largely vanishes for the metal-rich clusters: 28 clusters
are found within the major axis sectors compared to 26 along the minor
axis.  Based on their photometric metallicity estimates, \citet{els88}
discovered that their sample of clusters classified as metal-rich
defines a flattened system with an ellipticity of $\epsilon\gtsim 0.4$
within a major axis radius of $30\arcmin$.  Our cluster counts are not
consistent with such a large degree of flattening of the metal-rich
system in this region.  At present, we cannot provide reasonable
limits on the flattening of the metal-poor cluster distribution, since
the arrangement of the WYFFOS fields has resulted in a selection bias
in favor of clusters in major axis sectors at radii beyond $\sim
30\arcmin$.


\subsection{Metallicity Gradient}

\begin{figure*}
\hspace*{\fill}\psfig{file=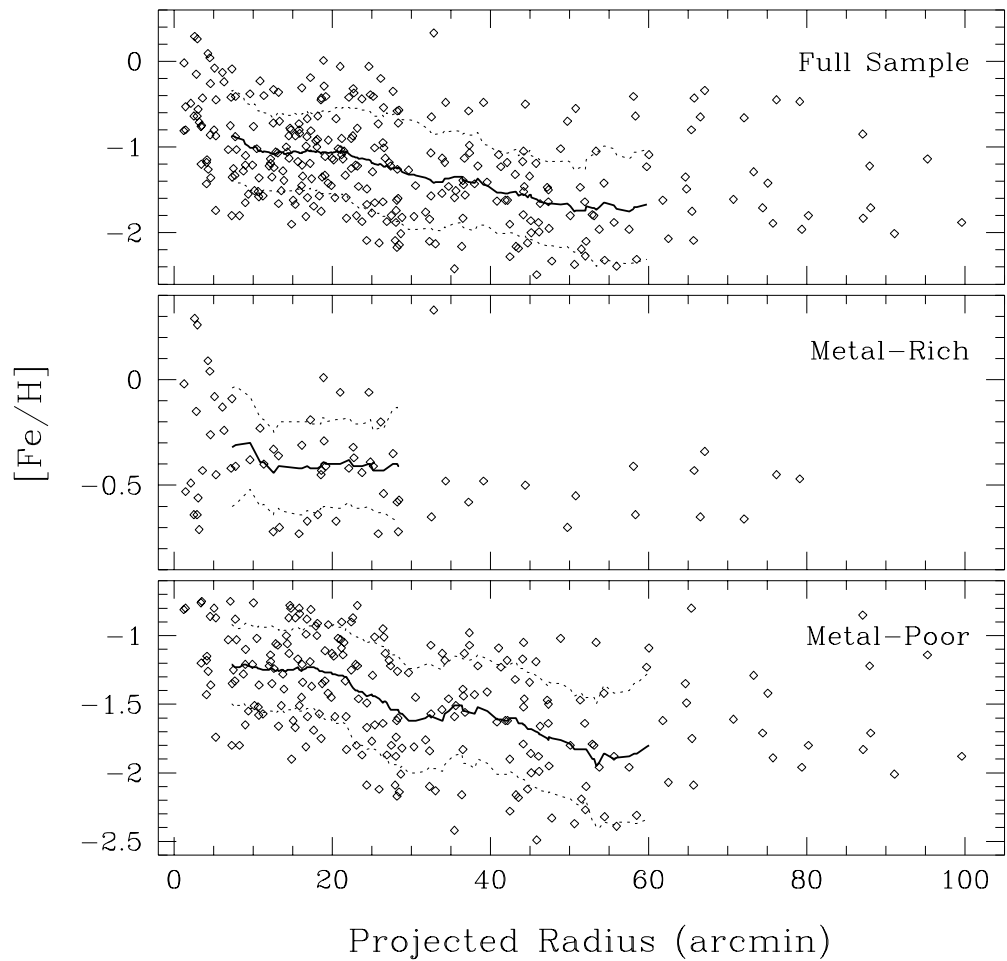,width=5.5in}
\hspace*{\fill}\caption{[Fe/H] as a function of projected radius for
the full sample and metallicity populations of the M31 GCS.  A
sliding-bin fit was used to determine the mean and RMS at each data
point (shown by the solid and dotted curves, respectively).}
\label{fig:grad}
\end{figure*}

Several previous works have suggested the presence of a weak
metallicity gradient in the M31 cluster system out to radii of
$\sim10$~kpc \citep{sha88, huc91}.  In Figure~\ref{fig:grad}, we plot
the metallicity of the M31 globular clusters as a function of
galactocentric radius for the full sample and the separate metallicity
sub-populations.  The large scatter in these plots is due at least in
part to the fact that we can only measure projected positions, but
nonetheless we find that the mean [Fe/H] does indeed decrease with
radius.

A sliding bin method was used to fit the mean metallicity in
$12\arcmin$ bins centered on each data point, with results shown in
Figure~\ref{fig:grad}.  Robust estimates of the mean metallicity in
radial bins are consistent with the presence of a modest gradient in
[Fe/H] within the full sample.  Separating the data into
sub-populations reveals no clear evidence for a gradient in the
metal-rich GCS, although the spread in metallicity does seem to
increase towards the center of the galaxy.  There is an apparent trend
towards decreasing metallicity within both the full sample and the
inner metal-poor cluster system, with gradients of roughly $-0.017$
and $-0.015~{\rm dex/arcmin}$, respectively.  Relatively poor sampling
of the clusters at large radii makes it difficult to comment
decisively on the results beyond a radius of about $60\arcmin$,
although the metal-poor sample appears to level off at a lower mean
metallicity beyond this distance.


\section{GCS Kinematics}
\label{sec:kinematics}

The maximum likelihood value for the mean cluster radial velocity is
$\langle{v}\rangle = -284\pm9$~\kms, somewhat higher than M31's
systemic velocity of $-300\pm4$~\kms\ (RC3).  The globular clusters
have an overall velocity dispersion of $\sigma_{v}=156\pm6$~\kms.  The
kinematics of metal-rich and metal-poor GCS populations will be
discussed later in this section.  First, we examine the global
kinematic properties of the M31 GCS.


\subsection{GCS Rotation}
\label{sec:rotation}

\begin{figure*}
\hspace*{\fill}\psfig{file=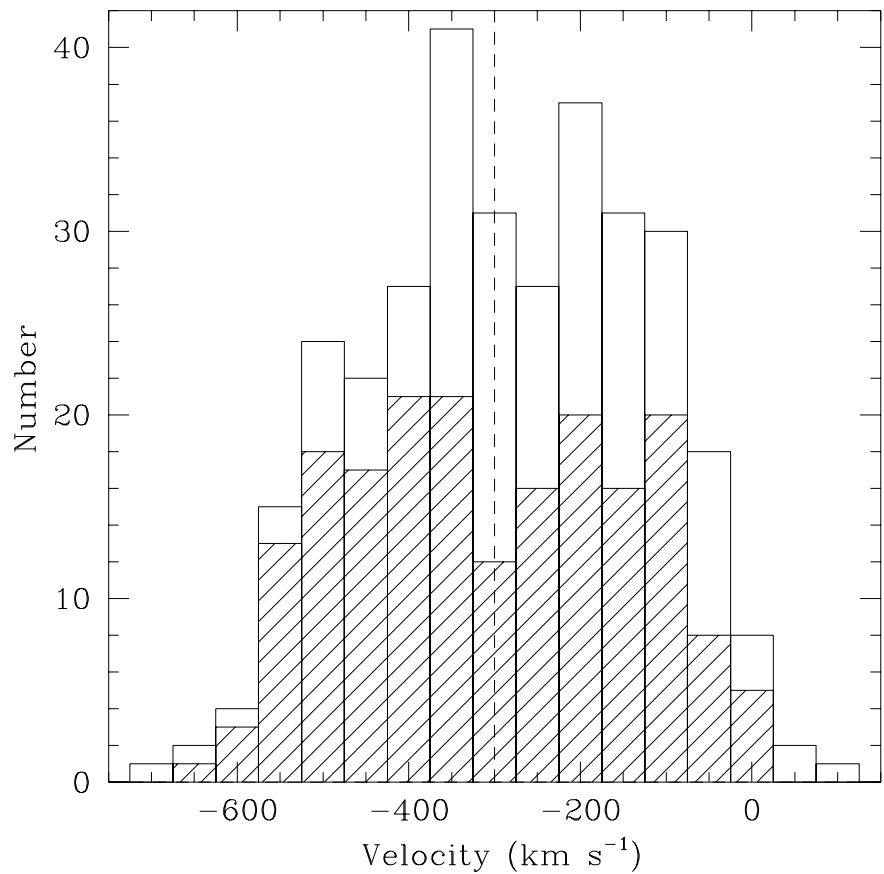,width=5in}
\hspace*{\fill}\caption{The velocity histogram for the M31 cluster
system.  The solid-line histogram is that for the available sample of
321 M31 globular cluster velocities, while the shaded area shows our
contribution of 191 WYFFOS velocities to the overall sample. The M31
systemic velocity of $-300 \pm 4$~\kms\ is shown by the dashed line.}
\label{fig:velhist}
\end{figure*}

A histogram of the cluster line-of-sight velocities is presented in
Figure~\ref{fig:velhist}.  There is strong evidence for bimodality in
the velocity distribution, a feature which is attributable to the fact
that the cluster system is rotating with respect to our line of sight.
Considering, for the moment, a purely rotating disk population of
globular clusters in the reference frame of the galaxy:
\begin{eqnarray}
v_{\rm GC}&=&v_{\rm rot}\sin{(i)}\sin{(\theta+\delta)}+v_\sigma  \\
          &=&v^\prime+v_\sigma  \nonumber
\end{eqnarray}
where $v_{\rm rot}$ is the rotation velocity of the population, $i$ is
the inclination of the rotation axis to the line of sight
($77.7^\circ$), $\theta$ is the angle about the angular momentum axis,
$\delta$ is an arbitrary phase offset, and $v_\sigma$ incorporates the
velocity dispersion plus random measurement errors.  The combined
probability distribution $P(v_{\rm GC})$ for the M31 globular cluster
radial velocities is the convolution of $P(v^\prime)$ with
$P(v_\sigma)$.  The probability distribution $P(v_\sigma)$ is Gaussian
and $P(\theta)$ is uniform, thus $P(v^\prime)$ is given by:
\begin{eqnarray}
P(v^\prime) &=& P(\theta)\frac{d\theta}{dv^\prime}\\
~ &=& \frac{1}{2\pi}\frac{1}{\sqrt{v^2_{\rm rot}
{\rm sin}^2{i}-{v^\prime}^2}}. \nonumber 
\end{eqnarray}
\noindent This represents a bimodal distribution with maxima at $\pm
v_{\rm rot}\sin{(i)}$ about the systemic velocity of the population.
The peaks of the (unshifted) velocity histogram in
Figure~\ref{fig:velhist} lie at $-415$ and $-170$~\kms, yielding an
estimated rotation amplitude of $v_{\rm rot}\sim 125$~\kms.

The fact that M31 is rotating has been established by various studies
\citep[see][and references therein]{hod92}.  Using a sample of
$\sim150$ M31 clusters as dynamical tracers, \citet{huc91} noted a
rapid rotation of the handful of inner ($R\ltsim 7\arcmin$) clusters
in their sample with ${\rm [Fe/H]}\geq -0.8$ (see also \citet{huc82}).
\citet{huc91} discovered that the metal-poor population had no
rotation signature within the inner galaxy, but shared a moderate
rotation of $\sim60$~\kms\ with the bulk of the GCS beyond this
radius.  With our increased sample of 321 velocities, we can now
provide a more comprehensive look at the kinematics of the M31
globular cluster system.

\begin{figure*}
\hspace*{\fill}\psfig{file=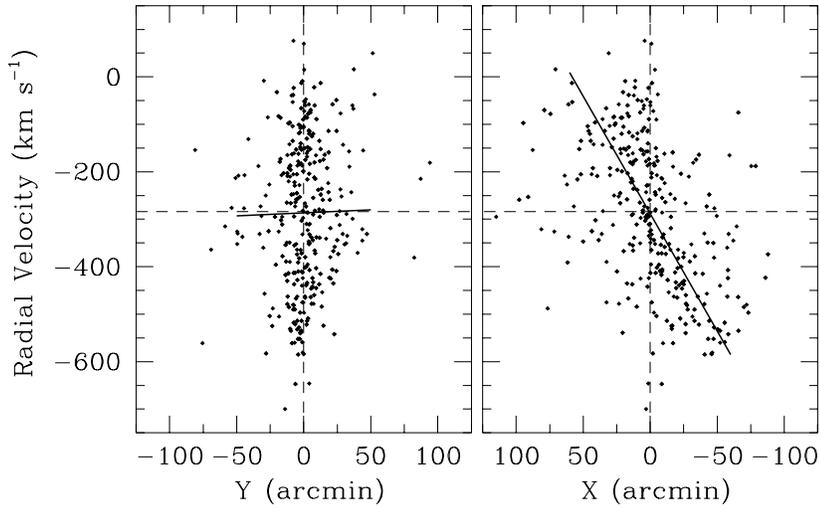,width=4.5in}
\hspace*{\fill}\caption{Radial velocity of the M31 globular clusters
versus projected radius along the minor axis (left) and the major axis
(right).  Culled robust fits to the data are shown by the solid
lines.}
\label{fig:velrot}
\end{figure*}

Perhaps not surprisingly, there appears to be no significant rotation
of the GCS about the major axis of M31 as shown in the left panel of
Figure~\ref{fig:velrot}.  A robust fit to the targets within the inner
$50\arcmin$ along the minor axis (dropping 11 objects at large
radii) gives $v_r = -286 + 0.13\,Y$~\kms, where $Y$ is the distance
along the minor axis in arcminutes.  Such a small deviation from a
slope of zero may simply be due to statistical imprecision or
variation in the position angle for the galaxy \citep{hod82}.

A plot of the observed cluster radial velocities as a function of
projected distance along the major axis of M31 is shown in the right
panel of Figure~\ref{fig:velrot}. A basic $2\sigma$ culled robust
linear fit to the velocities along the major axis reveals that the
globular cluster system indeed has a significant rotational component
about the minor axis.  This rotation will now be investigated in more
detail.

\begin{figure*}
\hspace*{\fill}\psfig{file=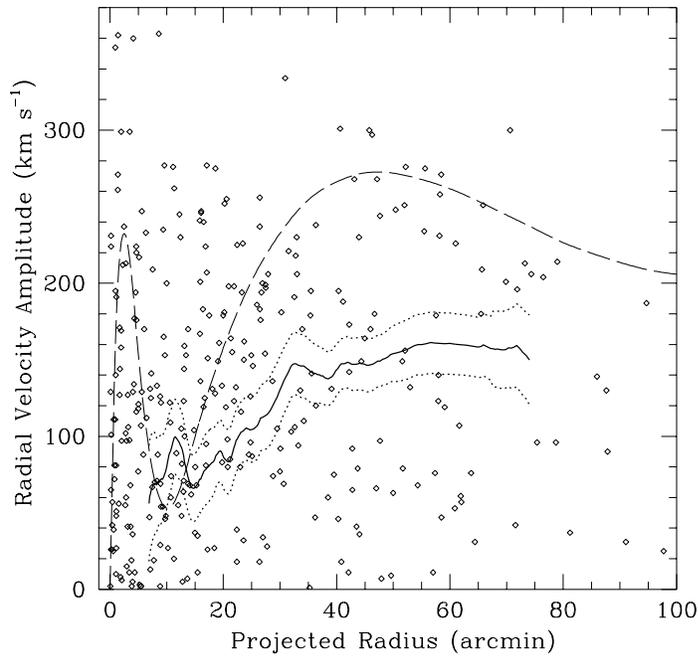,width=3.7in}
\hspace*{\fill}\caption{The projected rotation curve of the M31 GCS
and its 90\% confidence limits are shown by the solid and dotted
lines, respectively (see the text for details). The dashed line
represents the stellar rotation curve from \citet{rub70} for
comparison.}
\label{fig:rotfit}
\end{figure*}

We used an adaptive binning technique to generate the GCS rotation
curve shown in Figure~\ref{fig:rotfit}.  Before fitting the rotation
curve, the globular cluster coordinates were transformed into the
reference frame of the galaxy and the data were folded about each
axis.  The function $v=v_{\rm sys} + v_{\rm rot}\sin{\phi}$ was fit to
the radial velocities of the clusters within radial bins outwards from
the galaxy center \citep{kis98}.  The angle $\phi$ represents the
position angle of the target object in the projected frame of the
galaxy, and $v_{\rm sys}$ is the mean velocity of the cluster system.
The kinematic properties were determined in a series of annuli
centered on each cluster and encompassing a specified number of data
points (79 points, with a minimum of 39 at the extrema).  With this
method, we obtain a rotation amplitude of $136$~\kms\ where the curve
flattens off beyond $R=30\arcmin$.

\begin{figure*}
\hspace*{\fill}\psfig{file=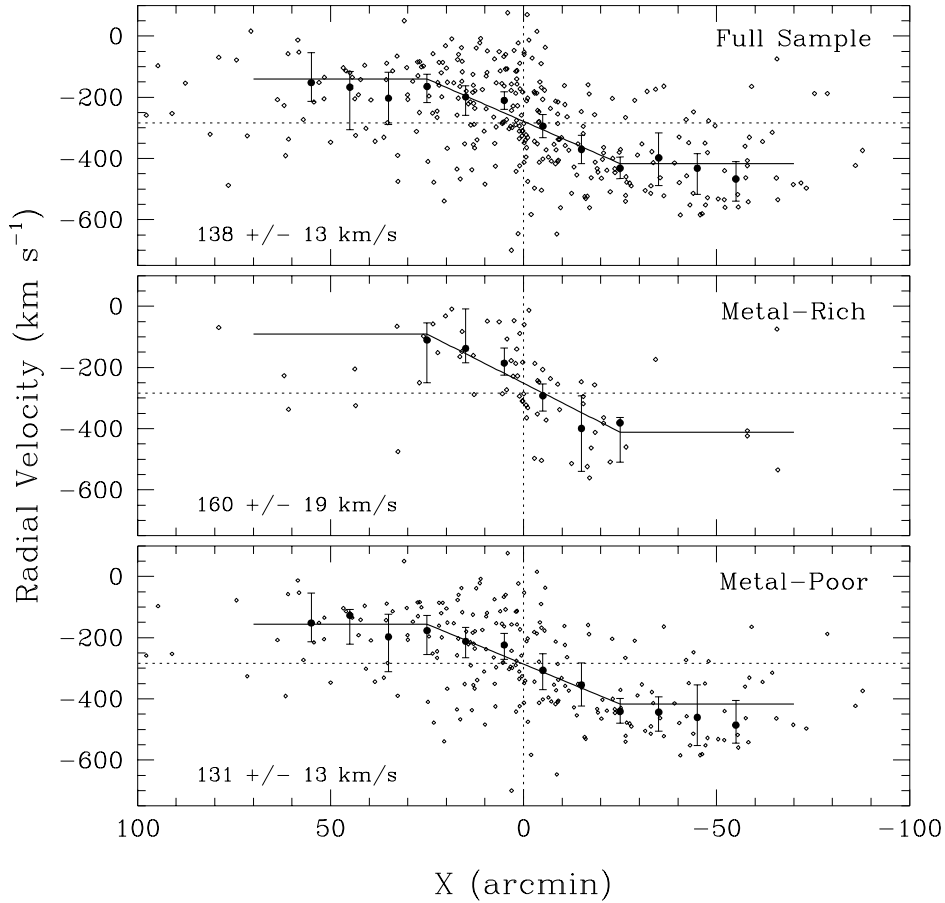,width=5in}
\hspace*{\fill}\caption{The rotation curve of the M31 GCS based on
robust biweight estimates of the mean (``location'') in radial bins of
10 arcminutes for the full dataset and separate metallicity
populations.  The error bars represent the 90\% confidence interval.
The solid line shows the best fit to the data assuming solid-body
rotation in the inner 25 arcminutes and flattening beyond.  The
derived rotation amplitudes are provided in the lower-left corner of
each plot.}
\label{fig:Rotation_bw}
\end{figure*}

We then employed the robust biweight estimators of location and scale
\citep{bee90, hoa00} as measures of mean velocity and dispersion in 10
arcminute bins along the major axis of the galaxy.  The biweight
results and their 90\% confidence intervals (based on bootstrap
errors) are presented in Figure~\ref{fig:Rotation_bw} for the full
sample and the two metallicity sub-populations.  Based on the
appearance of these plots, it seems reasonable to approximate the
shape of the curve as consisting of solid-body rotation within the
central $25\arcmin$, with a flattening beyond this radius.  One-half
of the velocity difference between the flat regions of the curve in
Figure~\ref{fig:Rotation_bw} yields a rotation amplitude of $v_{\rm
rot}=138\pm 13$~\kms\ for the full sample.

Significant differences between the kinematic properties of the
metal-rich and metal-poor clusters may indicate that these populations
underwent dissimilar formation mechanisms.  The maximum likelihood
value for the mean velocity of the 70 clusters in the metal-rich
sample is $\langle v \rangle = -260\pm18$~\kms\ with a dispersion of
$\sigma_{v} = 146\pm12$~\kms.  For comparison, the larger
metal-poor sample of 231 clusters demonstrates a somewhat lower mean
velocity of $\langle v \rangle = -290\pm10$~\kms\ and a slightly higher
dispersion of $\sigma_{v} = 155\pm7$~\kms, although these values
are still consistent with those of the higher metallicity sample
within the quoted errors.  From the fits of the biweight results, we
determine a rotation amplitude of $v_{\rm rot}=160\pm 19$~\kms\ for the
metal-rich population.  This value is somewhat larger than the
metal-poor amplitude of $v_{\rm rot}=131\pm 13$~\kms, although these
values are also not inconsistent within their formal errors.

\begin{figure*}
\hspace*{\fill}\psfig{file=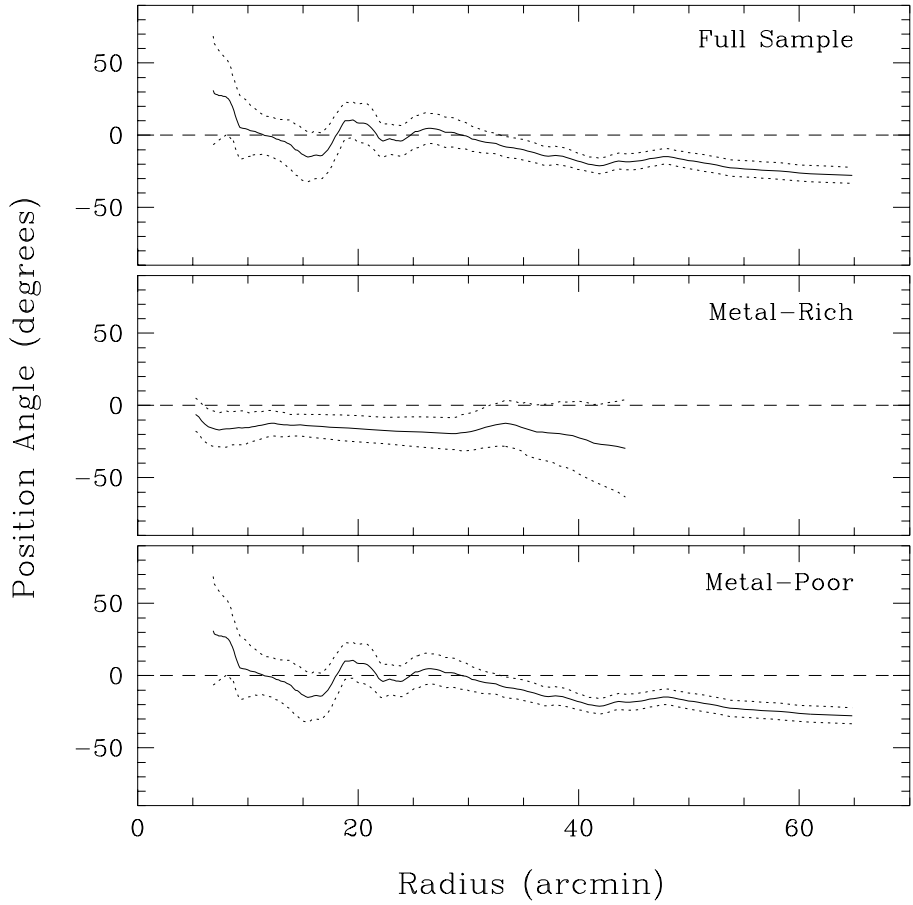,width=5in}
\hspace*{\fill}\caption{The projected position angle of the rotation
axis (solid line) and its 90\% confidence bands (dotted lines) as a
function of radius for the full sample, as well as for the metal-rich
and metal-poor populations.} 
\label{fig:PAplot}  
\end{figure*}

For each cluster, the projected position angle was determined as the
angle of maximum positive rotation obtained from the maximum
likelihood sinusoid fits \citep{kis98}. Plots of the variation of
position angle with radius for the full sample and metallicity
sub-populations are presented in Figure~\ref{fig:PAplot}. The input GC
positions already incorporate a de-rotation of ${\rm
PA_{M31}}=38^\circ$ for the galaxy and, all else being equal, one
would expect to find that the PA for the cluster system is roughly
zero.  Despite the sizeable spread at low radius in the full sample,
there is a relatively clear increase in PA within the central region
of the metal-poor cluster system.  This effect is consistent with
studies of the central bulge and inner disk region of M31 which
demonstrate a similar increase in PA \citep{hod82, ken89}.  The M31
GCS shows a notable decrease in the PA of the rotation axis beyond
about $30\arcmin$.  Interestingly, the metal-rich clusters seem to be
rotating about an axis that is tilted at a slightly lower PA with
respect to the major axis throughout the full radial extent of the
population.


\subsection{The Mass of M31}
\label{sec:mass}

The kinematics of M31's GCS provide us with a dynamical probe of the
underlying mass distribution of the host galaxy \citep{fed90, fed93}.
The projected mass estimator (PME) was used to place constraints on
the underlying mass of the system \citep{bah81,hei85}:

\begin{equation}
M_{\rm PME}=\frac{f_p}{NG}\sum_{i=1}^N r_i v_i^2,
\end{equation}

\noindent where $N$ is the number of bodies, G is the gravitational
constant, $r$ is the projected galactocentric radius and $v$ is the
radial velocity of the body relative to the systemic velocity of the
galaxy.  The factor $f_p$ depends on the assumptions made regarding
the distribution of orbits, and is equal to ${32}/{\pi}$ and
${64}/{\pi}$ for isotropic and radial orbits, respectively.

Admittedly, this projected mass estimate is hampered by our limited
sample size and coverage, as well as a lack of detailed understanding
of the cluster orbits.  Recent work by \citet{eva00} has shown that
the dynamical tracers in the M31 halo exhibit a predominantly
isotropic velocity dispersion.  Applying the PME to the available
cluster data and assuming isotropic orbits within an extended mass
distribution, we estimate a mass of $M_{\rm
M31}=(4.1\pm0.1)\times10^{11}~M_\odot$ using 319 dynamical targets out
to a radius of $\sim 120\arcmin$ from the galaxy center ($\sim 27$~kpc
at a distance of 770~kpc).  This result is comparable to previous mass
estimates obtained out to similar radii as summarized in Table~6 of
\citet{eva00}, and to the mass of $M_{\rm
M31}\sim(3.1\pm0.5)\times10^{11}~M_\odot$ found by \citet{huc91} based
on their sample of 150 globular clusters.  In the (unlikely) case of
purely radial orbits, our mass estimate would be increased to twice
the value quoted above.


\section{Discussion and Conclusions}
\label{sec:conclusion}

We have presented a sample of new spectroscopic observations of M31
globular clusters, contributing $\sim 200$ high-precision velocities
and [Fe/H] values for what represents a large fraction of the
dynamical tracers of the galaxy.  A comprehensive analysis of the
kinematics and abundance properties of M31's GCS demonstrates some
interesting similarities and differences when compared with the Milky
Way's globular cluster population.

The Galactic and M31 globular cluster systems have comparable mean
metallicities of ${\rm [Fe/H]}\sim -1.2$ to $-1.3$, although a
Kolmogorov-Smirnov test provides no conclusive evidence that the two
distributions are drawn from the same parent population.  A KMM
mixture-modeling test reveals that the M31 GCS metallicity
distribution can convincingly be separated into two groups, a
bimodality which is analogous to that found in the Milky Way GCS.  The
peaks in the M31 metallicity distribution lie at ${\rm [Fe/H]}=-1.44$
and $-0.50$.  A KMM test on the Milky Way GCS metallicities from the
McMaster catalogue \citep{har96} returns peaks at ${\rm [Fe/H]}=-1.59$
and $-0.56$, indicating that the metal-poor Galactic globulars have,
on average, a somewhat lower metal abundance than their M31
counterparts.  Of the 301 M31 globular clusters with available
metallicities, KMM assigns 70 to the metal-rich peak.  This represents
a somewhat smaller fraction of high-metallicity clusters than is found
in the Milky Way system, for which 45 of 145 are identified with the
metal-rich population.

M31's metal-rich globular clusters appear to constitute a distinct
kinematic subsystem that is quite spatially concentrated, consistent
with membership in a bulge population akin to the inner metal-rich
clusters of the Milky Way \citep{min95}.  Furthermore, the velocity
dispersion of the metal-rich cluster system, $\sigma_{v} =
146\pm12$~\kms, is indistinguishable from the mean bulge dispersion of
$\sigma_{v}{\rm (bulge)} = 146\pm6$~\kms\ as determined from the
kinematics of faint planetary nebulae \citep{law83, vdb91}. The mean
metallicity of this population is quite similar to their Milky Way
counterparts and demonstrates no obvious abundance gradient with
galactocentric radius.  Contrary to observations in the Milky Way,
however, we find no clear signs of flattening in the metal-rich
component of the M31 GCS despite our relative oversampling of major
axis fields in the current spectroscopic sample.  We find that the
rotation axis of the metal-rich cluster population is tilted at a
slightly lower position angle: it is offset by about $5-10^\circ$ with
respect to that of the minor axis of M31 and the bulk of its GCS.
This observation may point to a disparate origin for these clusters.
Further searches for potential cluster candidates hidden by the bulge
of M31 will be helpful in providing a clearer understanding of the
kinematics and nature of this inner system.

In M31, there are more than three times the number of metal-poor
clusters as there are metal-rich.  It is not clear if it is possible
(or even necessary) to further separate this larger, more spatially
distended metal-poor population into sub-components.  If the
metal-poor cluster distribution does indeed incorporate a thick disk
component in addition to halo clusters, this would account for the
relatively high net rotation amplitude and large velocity dispersion
observed therein.  While grouping the clusters based on metallicity
remains an instructive exercise, [Fe/H] alone is not an adequate
discriminator of membership in the spatially- or kinematically-defined
components of the galaxy. This is primarily a result of the
substantial spread and measurable gradient in [Fe/H] within the
metal-poor cluster system.  The task of assigning globular clusters to
their appropriate component (e.g. bulge, thin or thick disk, inner or
outer halo) remains somewhat simpler within the Milky Way GCS, since
here we have three-dimensional spatial information rather than
projected positions.

We find evidence for a radial metallicity gradient ($-0.015~{\rm
dex/arcmin}$) in the metal-poor population of M31 out to about
$60\arcmin$ from the galaxy center.  Despite a notable lack of cluster
data beyond this radius, it does seem that mean metallicity levels
off, although the scatter remains large.  This gradient in the
metal-poor clusters is not inconsistent with a single collapse
scenario, yet we recognize that a metallicity gradient, in and of
itself, is not sufficient to confirm an early dissipational collapse.
Such a gradient could be mimicked by a hierarchical halo formation if
the metallicity of the accreting fragments correlates with their mass
and mean density \citep{fre96}.  The search for substructure in
position/velocity/metallicity parameter space is clearly of interest
to investigate the accretion history of the galaxy.  An analysis of
potential sub-clustering within the M31 GCS will follow in a
subsequent paper.

A comprehensive investigation of the kinematics of the dynamical
tracers in M31 would not be complete without the addition of planetary
nebulae data.  Fortunately, a study of a large sample of M31 is
currently underway (Halliday et al., in preparation.  Moreover, these
data can be combined with the globular cluster sample to better probe
the underlying mass distribution of the galaxy (Wilkinson et al., in
preparation).  An analysis of M31 cluster ages is also in progress
(T.~Bridges, private communication) which should provide a better
sense of the chronology of GC formation within the galaxy.  Despite
its long and rich history of investigation, there is little doubt that
there are missing globulars in M31's database which are awaiting
discovery and identification.  Continuing searches and surveys for M31
cluster candidates \citep{lee01} promise to help remedy this situation
and to further fuel this ``booming industry''.  It is clear that
detailed studies of the kinematics, metallicities and ages of the
globular cluster systems of a large sample of galaxies can provide us
with the clues necessary for a more complete understanding of the
mechanisms which form and evolve galaxies of all kinds.


\section{Acknowledgments}

This work was supported in part by operating grants to DAH from the
Natural Sciences and Engineering Research Council of Canada.  JPB and
JPH acknowledge funding support from the National Science Foundation.
The authors wish to extend their thanks to Jim Lewis, Karl Gebhardt
and Tim Beers for making their software available, and to Pauline
Barmby for providing unpublished reddenings.  The authors are also
very grateful to the anonymous referee for the helpful comments and to
the kind staff at WHT for their support during the observations.



\clearpage

\begin{deluxetable}{cc}
\tablewidth{0pt}
\tablecaption{Target catalogue references\label{tab:targref}}
\tablehead{
 	\colhead{Code} & \colhead{Reference}
}
\startdata
B   & Battistini et al.~(1980, 1987)\\     
S   & Sargent et al.~(1977) \\             
BA  & Baade \& Arp (1964) \\               
BoD & Battistini et al.~(1987) Table VI \\ 
DAO & Crampton et al.~(1985) \\             
NB  & Battistini et al.~(1993) \\          
V   & Vete\u{s}nik (1962) \\               
\enddata
\end{deluxetable}

\begin{deluxetable}{ccccccc}
\tablewidth{0pt}
\tablecaption{Observing log\label{tab:obslog}}
\tablehead{
\colhead{Field ID} & 
\colhead{$\alpha_{\rm B1950}$} & 
\colhead{$\delta_{\rm B1950}$} & 
\colhead{Night} &
\colhead{UT} &
\colhead{Exp (Sec)} & 
\colhead{Grating} 
}
\startdata
Central 1 & 00:40:00.18 & +40:59:59.9 & Nov. 3/4 & 21:00 & 6 $\times$ 1800 
  & H2400B \\ 
Central 1 & 00:40:00.43 & +41:00:00.5 & Nov. 3/4 & 00:43 & 3 $\times$ 1200  
  & R1200R \\ 
Northeast 1 & 00:42:29.88 & +41:36:02.0 & Nov. 4/5 & 21:52 & 6 $\times$ 1800 
  & H2400B \\ 
Northeast 1 & 00:42:29.88 & +41:36:01.1 & Nov. 4/5 & 01:22 & 3 $\times$ 1200 
  & R1200R \\ 
Southwest 1 & 00:37:30.15 & +40:24:01.7 & Nov. 5/6 & 19:32 & 3 $\times$ 1200 
  & R1200R \\ 
Southwest 1 & 00:37:30.51 & +40:24:00.0 & Nov. 5/6 & 20:42 & 4 $\times$ 1800 
  & H2400B \\ 
Northeast 2 & 00:44:59.97 & +42:11:53.4 & Nov. 5/6 & 23:43 & 5 $\times$ 1800 
  & H2400B \\ 
Northeast 2 & 00:45:00.70 & +42:11:56.3 & Nov. 5/6 & 02:40 & 3 $\times$ 1200 
  & R1200R \\ 
Central 2 & 00:39:59.20 & +41:00:05.2 & Nov. 6/7 & 19:26 & 3 $\times$ 1200
  & R1200R \\ 
Central 2 & 00:39:59.34 & +41:00:03.7 & Nov. 6/7 & 20:38 & 4 $\times$ 1800 
  & H2400B \\ 
Southwest 2 & 00:34:59.80 & +39:47:54.6 & Nov. 6/7 & 23:33 & 5 $\times$ 1800 
  & H2400B \\ 
Southwest 2 & 00:35:00.56 & +39:47:54.6 & Nov. 6/7 & 02:25 & 4 $\times$ 1200 
  & R1200R \\ 
\enddata
\end{deluxetable}

\begin{deluxetable}{rccrrccl}
\tabletypesize{\footnotesize}
\tablewidth{0pt}
\tablecaption{WYFFOS results\label{tab:WHT}}
\tablehead{
\colhead{GC} & \colhead{RA (B1950)} & \colhead{Dec (B1950)} & \colhead{X (\arcmin)} & \colhead{Y (\arcmin)} & 
         \colhead{$v$ (km\,s$^{-1}$)} & \colhead{[Fe/H]} & \colhead{Notes}
}
\startdata
\multicolumn{8}{c}{(Complete table available in published version)}\\
\enddata                
\end{deluxetable}       

\begin{deluxetable}{lccc}
\tablewidth{0pt}
\tablecaption{Line Indices and Colors\label{tab:LIs}}
\tablehead{
 	\colhead{Index} & \colhead{C1} & \colhead{I} & \colhead{C2}
}
\startdata
HK          & 3910.00 -- 3925.00 &  3925.00 -- 3995.00 &  3995.00 -- 4015.00 \\
CNR         & 4082.00 -- 4118.50 &  4144.00 -- 4177.50 &  4246.00 -- 4284.75 \\
CaI         & 4200.00 -- 4215.00 &  4215.00 -- 4245.00 &  4245.00 -- 4260.00 \\
CH(G)       & 4268.25 -- 4283.25 &  4283.25 -- 4317.00 &  4320.75 -- 4335.75 \\
$\Delta$    & 3800.00 -- 4000.00 &      \nodata        &  4000.00 -- 4200.00 \\
H$\beta$    & 4829.50 -- 4848.25 &  4849.50 -- 4877.00 &  4878.25 -- 4892.00 \\
MgH         & 4897.00 -- 4958.25 &  5071.00 -- 5134.75 &  5303.00 -- 5366.75 \\
Mg2         & 4897.00 -- 4958.25 &  5156.00 -- 5197.25 &  5303.00 -- 5366.75 \\
MgG         & 5125.00 -- 5150.00 &  5150.00 -- 5195.00 &  5195.00 -- 5220.00 \\
Mgb         & 5144.50 -- 5162.00 &  5162.00 -- 5193.25 &  5193.25 -- 5207.00 \\
Fe52        & 5235.50 -- 5249.25 &  5248.00 -- 5286.75 &  5288.00 -- 5319.25 \\
Fe53        & 5307.25 -- 5317.25 &  5314.75 -- 5353.50 &  5356.00 -- 5364.75 
\enddata
\end{deluxetable}

\begin{deluxetable}{crrrrr}
\tablewidth{0pt}
\tablecaption{Correlation between line indices and [Fe/H]\label{tab:FeHfits}}
\tablehead{
 	\colhead{Line} & \colhead{$\sigma$} & \colhead{$a$} & 
	\colhead{$b$} & \colhead{RMS} & \colhead{$r$} 
}
\startdata
H\&K         & 0.10 & 0.51   & 0.16   & 0.12  & 0.37  \\ 
CNR          & 0.06 & 0.18   & 0.11   & 0.08  & 0.51  \\ 
CaI          & 0.06 & 0.00   & -0.01  & 0.06  & -0.05 \\ 
CH(G)$^\ast$ & 0.08 & 0.25   & 0.09   & 0.06  & 0.68  \\ 
$\Delta$     & 0.20 & 1.31   & 0.24   & 0.14  & 0.70  \\ 
H$\beta$     & 0.02 & 0.03   & -0.05  & 0.05  & -0.34 \\ 
MgH          & 0.04 & 0.13   & 0.02   & 0.04  & 0.52  \\ 
Mg2          & 0.05 & 0.20   & 0.11   & 0.05  & 0.72  \\ 
MgG          & 0.03 & 0.13   & 0.06   & 0.02  & 0.75  \\ 
Mgb$^\ast$   & 0.04 & 0.20   & 0.09   & 0.03  & 0.80  \\ 
Fe52         & 0.04 & 0.11   & 0.05   & 0.02  & 0.61  \\ 
Fe53$^\ast$  & 0.03 & 0.09   & 0.04   & 0.02  & 0.66  \\ 
\enddata
\tablecomments{Features marked with an asterisk were used in the final [Fe/H] 
calibration.}
\end{deluxetable}


\clearpage

\clearpage

\clearpage

\clearpage

\clearpage

\end{document}